\documentclass[AMA,STIX1COL]{WileyNJD-v2}

\usepackage{mathtools}
\usepackage{amsmath}
\usepackage{amsthm}

\usepackage[utf8]{inputenc}
\usepackage{graphicx}
\usepackage{pdfpages}
\usepackage{caption}
\usepackage{subcaption}
\usepackage{multirow}
\usepackage{tabularx}
\usepackage{float}
\usepackage{hyperref}
\usepackage{amsfonts}
\usepackage[export]{adjustbox}
\usepackage{csquotes}
\usepackage{listings}
\newcommand{\vc}[1]{\boldsymbol{#1}}

\newcommand{\reva}[1]{\textcolor{black}{#1}}
\newcommand{\revb}[1]{\textcolor{black}{#1}}
\newcommand{\revc}[1]{\textcolor{black}{#1}}

\articletype{Article Type}%

\received{}
\revised{}
\accepted{}

\begin{document}

\title{Automated generation of 0D and 1D reduced-order models of patient-specific blood flow}

\author[1,2,3]{Martin R. Pfaller*}
\author[4]{Jonathan Pham}
\author[4]{Aekaansh Verma}
\author[1,2]{Luca Pegolotti}
\author[5]{Nathan M. Wilson}
\author[6]{David W. Parker}
\author[1]{Weiguang Yang}
\author[1,2,3,7]{Alison L. Marsden}

\authormark{PFALLER \textsc{et al}}

\address[1]{\orgdiv{Pediatric Cardiology}, \orgname{Stanford University}, \orgaddress{\state{CA}, \country{USA}}}
\address[2]{\orgdiv{Institute for Computational and Mathematical Engineering}, \orgname{Stanford University}, \orgaddress{\state{CA}, \country{USA}}}
\address[3]{\orgdiv{Cardiovascular Institute}, \orgname{Stanford University}, \orgaddress{\state{CA}, \country{USA}}}
\address[4]{\orgdiv{Mechanical Engineering}, \orgname{Stanford University}, \orgaddress{\state{CA}, \country{USA}}}
\address[5]{\orgname{Open Source Medical Software Corporation}, \orgaddress{\state{CA}, \country{USA}}}
\address[6]{\orgdiv{Stanford Research Computing}, \orgname{Stanford University}, \orgaddress{\state{CA}, \country{USA}}}
\address[7]{\orgdiv{Bioengineering}, \orgname{Stanford University}, \orgaddress{\state{CA}, \country{USA}}}

\corres{*Martin R. Pfaller, 318 Campus Dr, Stanford, CA 94305. \email{pfaller@stanford.edu}}


\abstract[Summary]{Three-dimensional (3D) cardiovascular fluid dynamics simulations typically require hours to days of computing time on a high-performance computing cluster. One-dimensional (1D) and lumped-parameter zero-dimensional (0D) models show great promise for accurately predicting blood bulk flow and pressure waveforms with only a fraction of the cost. They can also accelerate uncertainty quantification, optimization, and design parameterization studies. \revb{Despite several prior studies generating 1D and 0D models and comparing them to 3D solutions, these were typically limited to either 1D or 0D and a singular category of vascular anatomies.} This work proposes a fully automated and openly available framework to generate and simulate 1D and 0D models from 3D patient-specific geometries\reva{, automatically detecting vessel junctions and stenosis segments}. Our only input is the 3D geometry; we do not use any prior knowledge from 3D simulations. All computational tools presented in this work are implemented in the open-source software platform \texttt{SimVascular}. We demonstrate the reduced-order approximation quality against \revb{rigid-wall} 3D solutions in a comprehensive comparison with $\revb{N=72}$ publicly available models from various anatomies, vessel types, and disease conditions. Relative average approximation errors of flows and pressures typically ranged from 1\% to 10\% for both 1D and 0D models, measured at the \revb{outlets of terminal vessel branches}.  \reva{In general, 0D model errors were only slightly higher than 1D model errors despite requiring only a third of the 1D runtime.} Automatically generated ROMs can significantly speed up model development and shift the computational load from high-performance machines to personal computers.}


\maketitle

\section{Introduction}

Image-based computational fluid dynamics (CFD) is increasingly used for patient-specific predictions of cardiovascular blood flow. Yet, three-dimensional (3D) methods typically require several hours of \revb{parallel computing.\cite{pfaller21}} Runtime is a severe limitation since applications of patient-specific modeling often require not only a single simulation, but numerous simulations. For example, iterative parameter estimation is often performed to match clinical targets, e.g., from \textit{in vivo} magnetic resonance flow imaging and catheter pressure measurements. Iterative optimization as well as parameter sweeps are often required to explore a design space with multiple parameters or to identify a personalized treatment plan.\cite{verma20b} In uncertainty quantification (UQ) or sensitivity analysis, several thousand simulations can be required to quantify the confidence in the simulation's predictions based on uncertainties in \revb{model} parameters.\cite{fleeter20,seo20} Finally, fast feedback is essential for clinical decision-making. A strategy to limit the computational demand is to employ reduced-order models (ROMs) whenever possible in the modeling pipeline, which can be run on a standard computer in seconds or minutes. While ROMs can accurately reproduce bulk flow and pressure waveforms they do not capture local flow features such as recirculation zones or local variations in wall shear stress. For widespread adoption, two requirements are necessary. First, the ROMs should be generated from the 3D models with minimal to no user interaction. Second, it is essential to quantify the accuracy of the ROMs against 3D solutions to assess their credibility in various anatomies and realistic conditions.

A variety of ROMs have been proposed in prior work. Reduced basis methods \cite{manzoni11} and proper orthogonal decomposition \cite{buoso19,pegolotti21} are standard tools to accelerate the solution of 3D CFD by utilizing results from previous solutions. \reva{Intermediate approaches between 3D and one-dimensional (1D) models exist by utilizing the pipe-like structure of blood vessels.\cite{alvarez16,alvarez19}} Similarly, machine learning methods use physics-informed neural networks to predict fluid dynamics.\revb{\cite{kissas20,fossan21}} In this work, we focus on zero-dimensional (0D) and 1D models, which are widely used and easily generalizeable to arbitrary anatomies. Analysis of the accuracy of 0D and 1D models across a large database of models will provide a baseline for future benchmarking of other ROM approaches.

\revb{Several previous studies have employed automatic generation of 1D or 0D models from 3D geometries. Stenoses have been detected using global minima and adjacent local maxima\cite{boileau17,blanco18} or by analyzing the slope of the vessel radius over the branch length.\cite{shahzad13,fossan18,mueller19} Vessel junctions have similarily been detected based on the local vessel radius.\cite{fossan18} While these models proved to be highly accurate compared to 3D models, their performance was only demonstrated for coronary arteries.}

LPNs have been widely used as stand-alone models and as boundary conditions in coupled 3D-0D simulations. Due to limitations in image resolution, the distal anatomy often cannot be included in the 3D model domain. Its influence is thus modeled in an LPN and coupled to the detailed 3D anatomical model. For example, 0D models can be used to represent the systemic, pulmonary, and coronary circulations, as well as the heart.\cite{migliavacca06,kim10a,kim10b,seo20b,gutierrez21} This approach was adopted to model the hemi-fontan surgery, where the circulatory system outside the 3D domain is represented by a 0D model.\cite{kung13}

Similar approaches have been used for 1D-0D coupled simulations, in which a higher fidelity 1D model is coupled to a lower fidelity 0D model, which provides the boundary conditions.\cite{olufsen99} A detailed 1D model of the human \revb{arterial and} venous system was coupled to a 0D model of the pulmonary circulation and heart chambers with valves.\cite{mueller14} A similar model was developed for the arterial tree.\cite{zhang15} Another study generated a network of 128 vessels to describe the human arterial system and compared it to blood pressure measurements.\cite{avolio80}

\revb{Parameter estimation under uncertainty is performed for automated boundary condition tuning in coupled 3D-0D models.}\cite{schiavazzi15,schiavazzi16b,tran17} A combination of 3D, 1D, and 0D models can be advantageous in multi-fidelity UQ approaches\reva{\cite{fossan18,fleeter20,guzzetti20}} and parameter estimation problems.\cite{seo20} Standard approaches for UQ in cardiovascular modeling pose challenges due to a large number of uncertain inputs and the high computational cost of realistic 3D simulations. Multilevel multifidelity Monte Carlo estimators improve the accuracy of hemodynamic quantities of interest while maintaining reasonable computational cost. This is achieved by leveraging three cardiovascular model fidelities, 3D, 1D, and 0D, each with varying spatial resolution, to quantify the variability in hemodynamic output. Note that for this application, it is not necessary that 1D and 0D models approximate the 3D solution with high accuracy. In fact, it is sufficient if 1D and 0D are reasonably correlated with 3D quantities of interest. A previous study reported good correlations in healthy and diseased models of aortic and coronary anatomy.\cite{fleeter20} However, it highlighted the need for a fully automated 1D and 0D modeling framework to facilitate the widespread use of UQ in cardiovascular simulations.

Several studies have quantified the approximation error of 1D or 0D models compared to high-fidelity 3D simulations. One study compared the outflow error in 0D vs. 3D simulations in 70 models of middle cerebral artery aneurysms during steady flow conditions.\cite{chnafa17} Outflow errors decreased significantly when considering energy losses at junctions.\cite{mynard15} Another study quantified the pressure drop across 22 mildly diseased human coronary arteries using a 0D model, taking into account curvature and stenosis under steady flow for different Reynolds numbers.\cite{schrauwen14} They found excellent agreement between 0D and 3D predicted pressure drops. It should be noted that the same data set was used first to fit the 0D model parameters, and then to validate the predictive capability of the model. Furthermore, only stenosed segments were considered, making it difficult to automate this approach for arbitrary blood vessels. Recently, a study compared 0d and 3D model predictions of pressure gradient in pulmonary artery stenosis models with good agreement.\cite{pewowaruk21} Here, the pressure drop was predicted purely from variations in the cross-sectional area.\cite{mirramezani20}

Good agreement was also found for pulsatile flow between 1D and 3D models for an idealized rigid-wall single vessel, a bifurcation, an aorta, and a patient-specific aorto-iliac artery and a porcine thoraco-thoraco aortic bypass.\cite{wan02} An in-depth comparison of 1D and 3D models for the aorta of a single patient was performed in another study, finding good reproduction of the pressure and flow waveforms.\cite{reymond13} \reva{For coronary arteries, a good match was obtained between estimates of the fractional flow reserve in 1D and 3D.\revb{\cite{boileau17,blanco18,mueller19}}} Two studies compared a 1D model of the Circle of Willis to 3D for a total of three patients.\cite{moore04,grinberg10} Good agreement was found between 3D and 1D solutions, albeit in one case only after manually tuning vessel resistances.\cite{moore04} Two further studies found good agreement between pressure and flow waveforms at multiple locations in several idealized \revb{compliant} arterial models, single vessel, bifurcation, aortic arch, and aorta.\cite{xiao13, hasan21} Several studies compared 1D solutions to a 3D whole arterial tree model.\cite{reymond09, reymond12, boileau15, bertaglia20, blanco20} In summary, there was reasonable agreement in larger vessels and healthy scenarios but less in anatomical variations or abnormal hemodynamic conditions.

Prior studies have validated results from 1D blood flow models against \emph{in vitro} experimental and \textit{in vivo} data. \revb{In \textit{in vivo} studies,} blood flow is commonly extracted from phase-contrast magnetic resonance imaging (PC-MRI). In a study of bypass grafts in stenosed porcine aortas \cite{steele03}, energy losses in stenoses and junctions were taken into account, although they required a manual extraction of minimal and maximal cross-sectional areas from imaging. This comparison produced less than 11\% error in the flow ratios of eight studied animals. Further studies compared the whole human arterial tree \cite{reymond09} and a network of major arteries and major veins,\cite{mueller14} revealing that arterial flow waveform patterns were in accordance with PC-MRI measurements with reasonable blood flow distribution. Other studies have also explored \textit{in vitro} experiments, whereby hydraulic replicas of blood vessels enable simultaneous measurement of local flows and pressures. For example, an \textit{in vitro} model of a human arterial tree with 37 branches driven by a pulsatile pump was compared at 70 locations to simulations using an elastic \cite{matthys07} and visco-elastic tube law.\cite{alastruey11} The updated version \cite{alastruey11} achieved root-mean-square errors of 2.5\% and 10.8\% for pressure and flow, respectively.

While good quality approximations of bulk flow and pressure can be obtained from 1D and 0D models, there is currently no openly available and fully automatic framework to generate these models, limiting their applicability and adoption by the community. Furthermore, previous comparisons to high-fidelity 3D CFD were usually limited to specific vessels and healthy subjects or specific disease types in a few geometries. Our goal in this work is thus twofold. First, we propose a fully automated framework to generate 1D and 0D ROMs from 3D vascular geometries. This framework does not require any user-interaction and is openly available in \texttt{SimVascular} (\url{http://simvascular.org}).\cite{updegrove16} Second, we demonstrate the robustness of our framework and quantify the approximation quality of 1D and 0D models by comparing them to $\revb{N=72}$ high-fidelity rigid-wall 3D solutions from the Vascular Model Repository.\cite{wilson13} We compare model errors at the inlets, outlets, and interiors of the models for a large variety of patient anatomies, vessel types, and diseased states under pulsatile flow conditions. \reva{We additionally show studies of branch-refinement and the application of our framework to deformable wall simulations.}

\section{Materials and Methods \label{sec_methods}}
In this section, we briefly review the methods for 3D, 1D, and 0D models. A comprehensive derivation of all three model fidelities is provided elsewhere.\cite{quarteroni16} Furthermore, we introduce our automated ROM pipeline, where we extract all necessary information from the 3D geometry. 

\subsection{3D modeling \label{sec_theory_3d}}
Blood flow in the cardiovascular system is \revb{generally modeled by the incompressible} Navier-Stokes equations, 
\begin{align}
\rho \left( \dot{\vc{u}} + \vc{u} \cdot \nabla \vc{u} \right) &= \nabla \cdot \vc{\tau} + \rho \vc{b}, \quad \nabla \cdot \vc{\tau} = -p\vc{I} + \mu\left( \nabla \vc{u} + \nabla\vc{u}^\intercal \right), & \vc{x} \in \Omega^{3D},~t \in \mathbb{R}_{\geq 0},\label{eq_3d_momentum}\\
\nabla \cdot \vc{u} &= 0, & \vc{x} \in \Omega^{3D},~t \in \mathbb{R}_{\geq 0},\label{eq_3d_mass}
\end{align}
a set of three-dimensional partial differential equations, describing the relationship between the velocity field $\vc{u}\left(\vc{x}, t\right)$ and the pressure field \revb{$p\left(\vc{x}, t\right)$} for blood with a density $\rho$ and a dynamic viscosity $\mu$, subject to a body force, $\vc{b}\left(\vc{x}, t\right)$. Here, $\vc{\tau}\left(\vc{x}, t\right)$ is the stress tensor, $\vc{I}$ is the identity matrix, and $\Omega^{3D}$ is the volume domain of the fluid. Equation~\eqref{eq_3d_momentum} is the differential form of Newton's second law, applied to fluids in an Eulerian framework. Equation~\eqref{eq_3d_mass} is the continuity equation \revb{for an incompressible fliud}, which simply states that mass is conserved in the fluid system.

These equations are typically solved numerically to obtain spatial and temporal distributions of hemodynamics, including the velocity and pressure, in computational models of patient-specific vascular anatomies. To close this system of equations, initial and boundary conditions must be specified. The initial conditions are 
\begin{align}
\vc{u}(\vc{x}, t = 0) &= \vc{u_0}(\vc{x}), & \vc{x} \in \Omega^{3D},\label{eq_3d_ic_u}\\
\quad p(\vc{x}, t = 0) &= p_0(\vc{x}), & \vc{x} \in \Omega^{3D}.\label{eq_3d_ic_p}
\end{align}
In patient-specific modeling, the entire cardiovascular system cannot be geometrically modeled. Rather, it is common to model just the anatomical portion of interest, such as the aorta. The locations where the model ends define the inlet and outlet caps. \revb{To model the effects of blood vessels adjacent to the 3D domain $\Omega^{3D}$}, we prescribe boundary conditions at these cap surfaces. There are generally two types of boundary conditions applied here: open-loop and closed-loop conditions. In open-loop models, the inlet and outlet boundary conditions are applied separately and are not mathematically related. In closed-loop models, the inlet and outlet boundary conditions are numerically coupled. In this work, we focus on open-loop models for simplicity, though the methods we present are generalizable. We refer interested readers to further literature for details on closed-loop models. \revb{\cite{moghadam12,sankaran12,mueller14,toro21}}

A common open-loop inlet boundary condition prescribes a flow rate, $Q_\text{in}$, with a given velocity profile, $\vc{u}_\text{in} (\vc{x}, t)$, normal to the inlet, commonly using a parabolic profile.\cite{taylor98} This Dirichlet boundary condition is described by
\begin{align}
\vc{u}(\vc{x}, t) &= \vc{u}_\text{in}(t), & \vc{x} \in \Gamma_{in},~t \in \mathbb{R}_{\geq 0}, \label{eq_3d_in}
\end{align}
where $\Gamma_{in}$ represents the inlet cap surface of the model. At the outlets, resistance and Windkessel boundary conditions, capturing the viscous and compliant nature of downstream vessels, are commonly employed in open-loop models.\cite{vignonclementel06} These boundary conditions generally relate the flow rate to the pressure via algebraic-differential equations, as described by 
\begin{align}
P_\text{out}(\vc{x}, t)&= f(\vc{x}, t, Q_\text{out}(t), \dot{Q}_\text{out}(t), \vc{\phi}), & \vc{x} \in \Gamma_\text{out},~t \in \mathbb{R}_{\geq 0}, \label{eq_3d_out}
\end{align}
where $\vc{\phi}$ is a set of parameters governing the lumped-parameter elements and $\Gamma_\text{out}$ is an outlet cap surface. The values of these parameters are patient-specific and generally require tuning to match clinical targets.\cite{tran17} Finally, at the walls of the blood vessels, no-slip conditions,
\begin{align}
\vc{u}\left(\vc{x}, t\right) &= \vc{0}, & \vc{x} \in \revb{\Gamma_w},~t \in \mathbb{R}_{\geq 0},
\label{eq_no_slip}
\end{align}
are typically assumed, where $\Gamma_{w}$ is the wall surface domain of the fluid.

In this work, we created three-dimensional patient-specific vascular models using \texttt{SimVascular}, an open-source software providing a full pipeline for cardiovascular model generation and simulation.\cite{updegrove16} \reva{Note that we consider only rigid-wall behavior due to the availability of 3D rigid wall simulations in a large data repository, though our methods can be generalized to the setting of fluid-structure-interaction as demonstrated in Section~\ref{sec_fsi}.}

We spatially discretize the Navier-Stoke equations using a P1-P1 finite element formulation stabilized via SUPG and PSPG and use the generalized-$\alpha$ method for time advancement.\cite{franca92,taylor98,whiting01} We simultaneously solve the algebraic-differential equations governing the boundary conditions and the linear system resulting from finite element discretization of the Navier-Stokes equations using a modular implicit coupling scheme and a custom linear solver and preconditioner.\cite{moghadam13,esmailymoghadam13} Our \texttt{svSolver} finite element implementation of the 3D solver for cardiovascular flows is available open-source at \url{https://github.com/SimVascular/svSolver}. 

Additionally, running 3D simulations for cardiovascular models requires simulating multiple cardiac cycles to achieve results that have converged to a periodic state. This process can easily consume several days, even while using multiple processors and high-performance computing clusters. As such, we use \revb{our previously plublished} method to initialize our 3D simulations and minimize the number of cardiac cycles required for each patient-specific model considered in this work.\cite{pfaller21} All 3D simulations were run on Stanford's Sherlock supercomputing cluster using four 12-core Intel Xeon Gold 5118 CPUs. We ensured that the pressure error to the periodic state in the 3D solutions is below 1\% at all outlets \revb{using our 0D periodicity check}.\cite{pfaller21} \revb{Here, we use the repeated outflows of the 3D simulation during the last cardiac cycle and feed them into the LPN models of the boundary conditions. The 3D simulation has then reached a periodic state if the pressures are within 1\% of the converged LPN pressures.}

\subsection{1D modeling \label{sec_theory_1d}}
By integrating the Navier-Stokes equations \eqref{eq_3d_momentum} and \eqref{eq_3d_mass} over the lumen cross-section and assuming an axisymmetric velocity profile, we obtain the one-dimensional equations,
\begin{align}
\frac{\partial Q}{\partial t} + \frac{4}{3} \frac{\partial}{\partial z} \left(\frac{Q^{2}}{S}\right) + \frac{S}{\rho}\frac{\partial p }{\partial z} &= Sf -N\frac{Q}{S} + \frac{\mu}{\rho}\frac{\partial^2 Q}{\partial z^2}, & z \in \Omega^{1D}, t \in \mathbb{R}_{\geq 0} 
\label{eqn_1d_mom}, \\
\frac{\partial S}{\partial t} + \frac{\partial Q}{\partial z} &= 0, & z \in \Omega^{1D},~t \in \mathbb{R}_{\geq 0}.
\label{eqn_1d_mass}
\end{align}  
which \revb{govern} the interaction between the flow rate $Q\left(z, t\right)$, pressure $P\left(z, t\right)$, and cross-sectional area $S\left(z, t\right)$, subject to a body force $f\left(z, t\right)$, along the blood vessel's centerline axial coordinate, $z$.\cite{hughes73} Here, $\Omega^{1D}$ is the centerline domain of the blood vessel. Observe that unlike the 3D equations, the 1D equations only yield the temporal and axial distributions of bulk hemodynamic quantities. \reva{The variable $N$ is determined by the choice of velocity profile. For a quadratic, i.e., Poiseuille, flow profile, the expression becomes}
\begin{align}
\reva{N=\frac{8\pi\mu}{\rho}.}
\label{eqn_1d_n}
\end{align}  
An additional constitutive relationship between the pressure and the cross-sectional area is required to close the system. In this work, we use both \reva{a linear and nonlinear (Olufsen)} constitutive material model,
\begin{align}
\text{linear:~} P(z,t)=P^0(z)+k_0 \left( \sqrt{\frac{S(z,t)}{S^0(z)}} - 1† \right), \quad
\text{Olufsen:~} P(z,t)=P^0(z)+\frac{4}{3} \left(k_1e^{k_2r^0(z)}+k_3\right) \left( 1-\sqrt{\frac{S^0(z)}{S(z,t)}} \right),
\label{eqn_1d_mat}
\end{align}
\reva{where $E$ is the Young modulus of the material that composes the vessel wall}, $h$ is the wall thickness of the vessel, $P^0$ is a reference pressure, $r^0$ is a reference radius, and $\reva{k_0}, k_1$, $k_2$, and $k_3$ are empirically derived material constants.\cite{olufsen99} Observe that this system of equations enables us to simulate deformable wall behavior. However, for comparison against our 3D models in this work, we invoke rigid-wall behavior in our 1D models by setting $k_1$ to zero and $k_3$ to an arbitrarily large value. The initial conditions for the 1D system of equations are
\begin{align}
Q(z, t = 0) &= Q^0(z), & z \in \Omega^{1D},\label{eqn_1d_ic_q}\\
S(z, t = 0) &= S^0(z), & z \in \Omega^{1D}.\label{eqn_1d_ic_s}
\end{align}
Note that initial conditions for pressure do not need to be provided, given that pressure is directly related to the cross-sectional area via Equation \eqref{eqn_1d_mat}. At the outlets, we prescribe the same LPN boundary conditions used in 3D to represent the downstream vasculature.\cite{vignon04} The 1D Equations \eqref{eqn_1d_mom} and \eqref{eqn_1d_mass} govern only flow in the tubular sections of our patient-specific vascular networks. At the junctions between the different blood vessels in our model, we assume that \reva{static} pressure is continuous across the junction and mass is conserved. \reva{The Poiseuille flow assumption underestimates the pressure loss in stenoses. We incorporate pressure losses across stenoses via the heuristic formula \cite{wan02}}
\begin{align}
\reva{N = - \frac{S_s^2 Q_0^2 \left[ \frac{K_v}{\text{Re}} + \frac{K_t}{2} \left( \frac{S_0}{S_s} - 1 \right) \right]}{S_0^2Q_1 L}, \quad K_v = 32 \frac{L}{D_0} \left( \frac{S_0}{S_s} \right)^2, \quad K_t = 1.52,}
\label{eqn_1d_stenosis}
\end{align}
\reva{where subscript $0$ refers to the segment proximal to the stenosis, subscript $1$ refers to the stenosed segment. In addition, Re is the Reynolds number, $D_0$ is the cross-sectional diameter, and $L$ is the length of the stenosis. Note that $S_s=S_0$ recovers the Poiseuille solution \eqref{eqn_1d_n}, which we also use as a lower bound of $N$. We detail the selection of the parameters of the stenosis segment \eqref{eqn_1d_stenosis} in Section~\ref{sec_stenosis}.}

We numerically solve the 1D equations, coupled with the constraints at the junctions, using \revb{an implicit} Discontinuous Galerkin space-time finite element method with piecewise linear shape functions \revb{in space and piecewise constant shape functions in time}.\cite{wan02} Our open-source implementation of the 1D solver can be found at \url{https://github.com/SimVascular/svOneDSolver}. As with our 3D models, we generate the 1D patient-specific models using \texttt{SimVascular}.\cite{updegrove16} However, unlike 3D simulations, solving the 1D equations is computationally inexpensive and thus all 1D simulations were performed using only a single processor.

\subsection{0D modeling \label{sec_theory_0d}}
Zero-dimensional models are LPNs that simulate only bulk hemodynamic quantities, primarily flow rate and spatially averaged pressure, and their temporal distributions. These models are built from individual lumped-parameter elements. The primary building blocks for 0D models are resistors, capacitors, and inductors, analogous to electrical circuits. The flow rate in the 0D model corresponds to the current in an electrical circuit. The pressure drop across a 0D element mimics voltage drop. In the context of cardiovascular modeling, resistance models the viscous effects of the blood flow, capacitance models the elastic deformability of the blood vessel walls, and inductance models the inertia. The governing equations for these elements are 
\begin{align}
\Delta P &= RQ, &
Q &= C \Delta \dot{P}, &
\Delta P = L \dot{Q},
\end{align}
where $P$ is the pressure, $Q$ is the flow rate, $R$ is resistance, $C$ is capacitance, and $L$ is inductance. The values of the resistance, capacitance, and inductance are dependent on the cardiovascular anatomy. However, in general, for straight blood vessels under fully developed Poiseuille flow, these quantities can be computed from the viscosity of the blood, the \revb{linear} material properties of the vessel wall, and the geometry of the vessel.\cite{milisic04} \revb{The 0D elements are commonly described by}
\begin{align}
R &= \frac{8\mu l}{\pi r^4}, &
C &= \frac{3l\pi r^{3}}{2Eh} , &
L &= \frac{\rho l}{\pi r^{2}}.
\label{eqn_0d_mat}
\end{align}
Here, $\mu$ is dynamic viscosity of the blood, $\rho$ is the density of the blood, $r$ is the radius of the lumen, and $l$ is the longitudinal length of the blood vessel. 

Blood vessels with steep gradients in the cross-sectional area along the axial dimension, in particular, stenosed vessels, may experience flow separation effects, which render the Poiseuille-flow assumption invalid. A nonlinear expansion-based resistance,
\begin{align}
R_\text{expansion} &= K_{t}\frac{\rho}{2\reva{S_{0}^{2}}}\left(\reva{\frac{S_{0}}{S_{s}}} - 1\right)^{2}|Q|,
\label{eqn_0d_stenosis}
\end{align}
can be augmented to the Poiseuille resistance in the 0D model to account for such separation effects.\cite{mirramezani19,mirramezani20} Here, $K_{t}=1.52$ is a commonly used empirical correction factor.\cite{steele03,itu12,mirramezani19} The \reva{areas $S_{0}$ and $S_{s}$} are the cross-sectional areas of the lumen proximal to and at the location of the stenosis. \reva{Note that the 1D pressure drop modeled by \eqref{eqn_1d_stenosis} is equivalent to the 0D pressure drop in \eqref{eqn_0d_stenosis}.} \reva{As in 1D, we assume that static pressure is continuous across junctions and mass is conserved.}

The 0D simulation methods discussed above have been implemented as an open-source Python package available at \url{https://github.com/SimVascular/svZeroDSolver}. \texttt{svZeroDSolver} incorporates the building-block-like nature of 0D models into a highly modular software package for constructing and simulating arbitrary 0D models. A variety of commonly used 0D elements, such as a Poiseuille-based resistor, are implemented. Associated with each 0D building block are the equations \revb{governing the 0D flow physics in that element}. The local building blocks, along with these local equations, are assembled to construct the full 0D model, which yields a global system of differential-algebraic equations governing the entire 0D model:\cite{ascher98,verma20}
\begin{align}
\vc{E}\left(\vc{y}, t\right)\cdot\dot{\vc{y}} + \vc{F}\left(\vc{y}, t\right)\cdot \vc{y} + \vc{c}\left(\vc{y}, t\right) = \vc{0}.
\label{eqn_0d_dae}
\end{align}
Here, $\vc{y}$ is the global vector of solution variables, including the flow rate and pressure for each 0D element, $\vc{E}$ and $\vc{F}$ are their associated coefficients, and $\vc{c}$ is a vector of constants. Note that $\vc{E}$, $\vc{F}$, and $\vc{c}$ could be functions of $\vc{y}$, as in the case of the nonlinear expansion-based resistance in Equation~\eqref{eqn_0d_stenosis}.  We advance Equation~\eqref{eqn_0d_dae} in time using the implicit generalized-$\alpha$ method,\cite{jansen00} with the Newton-Raphson method for linearization, to simulate the hemodynamics in our 0D models. A deeper discussion of the mathematical details and implementation of our \texttt{svZeroDSolver} software can be found in Appendix~\ref{sec_appendix_0d}. Furthermore, similar to the 1D models, we apply the same boundary conditions to our 0D models as used in our 3D models. Finally, we generate 0D models using \texttt{SimVascular} \cite{updegrove16} and perform the 0D simulations using only a single processor.

\subsection{Automated ROM generation \label{sec_framework}}
In this section, we outline the steps to automatically generate reduced-order 1D and 0D models from 3D patient-specific geometries. The automated ROM generation methods described in this section have been incorporated into the \texttt{2021.09.30} release (\url{https://simtk.org/projects/simvascular}) of the open-source \texttt{SimVascular} application.\cite{updegrove16} The \texttt{SimVascular} ROM Simulation Tool (\url{https://simvascular.github.io/docsROMSimulation.html}) provides a graphical user interface for interactively setting parameters that control how a ROM is generated. Models can also be generated progamatically using the \texttt{SimVascular} Python Interface (\url{https://simvascular.github.io/docsPythonInterface.html}). This allows one to integrate 1D and 0D model generation and simulation into user-defined workflows and scripts, e.g. for parameter estimation or uncertainty quantification.

\subsubsection{Centerline extraction \label{sec_centerline}}
We automatically compute the centerlines with the help of the Vascular Modelling Toolkit (VMTK)\revb{, using the \texttt{vtkvmtkPolyDataCenterlines} and \texttt{vtkvmtkPolyDataCenterlineSections} classes with default arguments and custom modifications detailed below.}\cite{antiga08} The centerlines are defined as lines between the inlet and an outlet, whose minimal distance from the surface is maximized in a suitable norm.\cite{piccinelli09} Each point on the centerline is assigned a maximum inscribed sphere radius (MISR). Each individual centerline connects the inlet to one outlet. Points are created for each centerline and connected to form a path segment. We use the Visualization Toolkit (VTK)\cite{schroeder00} to merge all individual centerlines and remove duplicate points. This greatly reduces the number of centerline points and avoids redundant points, especially for geometries with many outlets. We further apply light global smoothing and moderate local smoothing close to the caps using a moving average smoothing filter.


\subsubsection{Cross-sectional area \label{seg_area}}
\begin{figure}[Htp!]
\centering
\includegraphics[height=5.3cm]{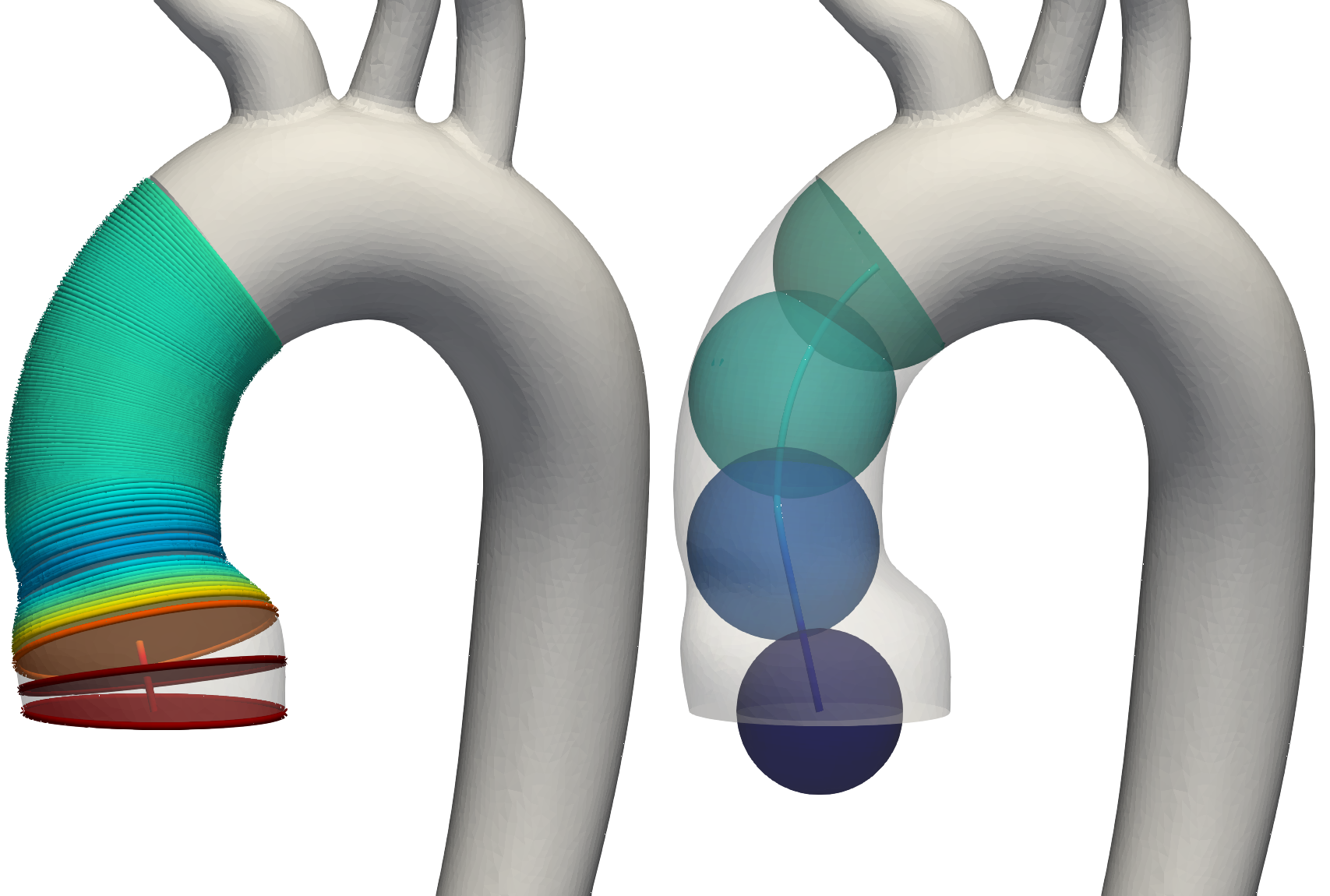}
\includegraphics[height=5.3cm]{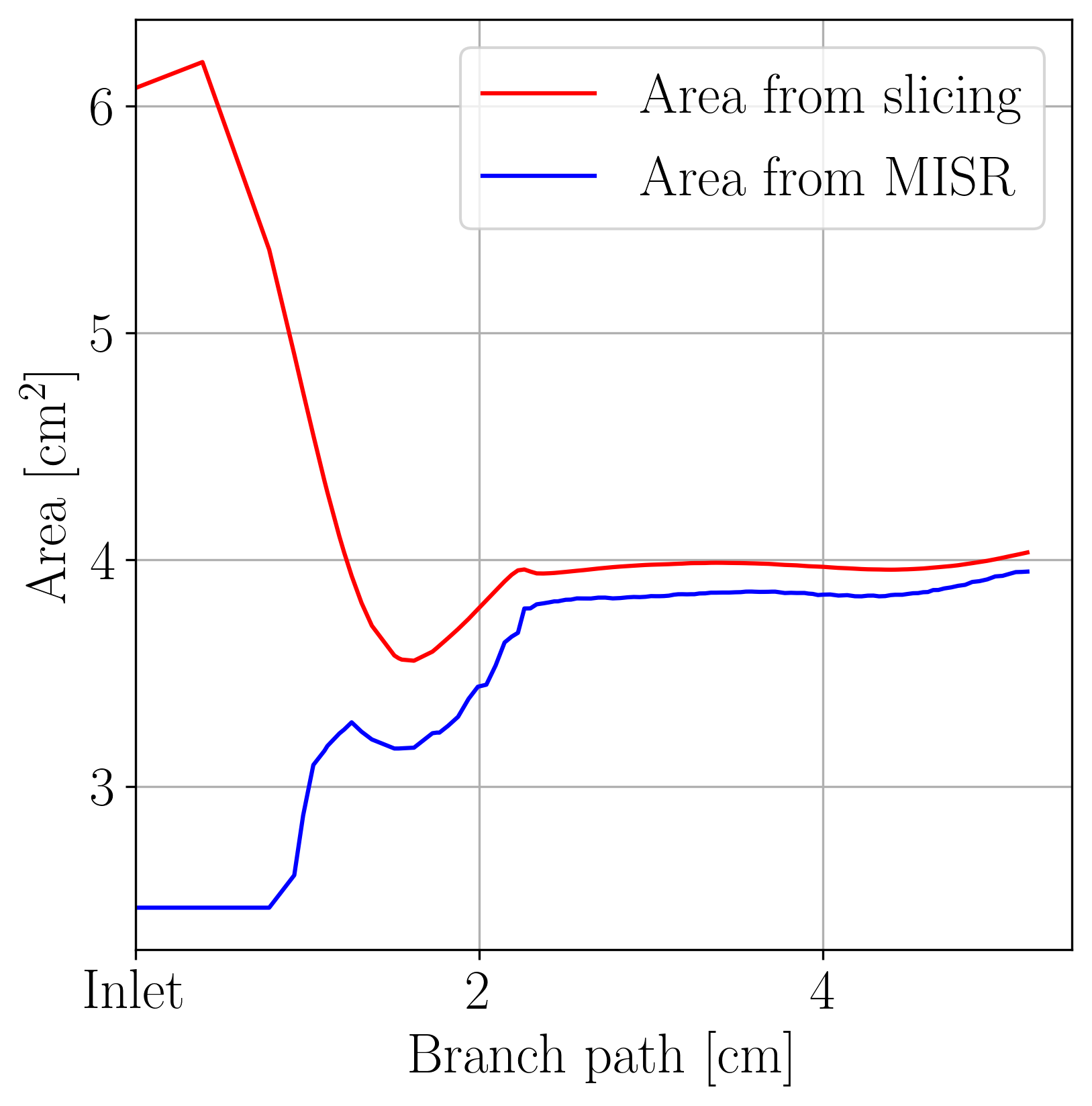}
\caption{Computation of cross-sectional area using slicing (left) and maximum inscribed sphere radius (MISR, middle). The color shows cross-sectional area in both images. The graph (right) shows both measures of cross-sectional area over the path of the branch, starting at the inlet. \label{fig_radius_vs_slice}}
\end{figure}

The cross-sectional area of the vessels is an important parameter in reduced-dimensional modeling. The area is the only geometrical measure besides the vessel segment length entering the reduced models. VMTK provides two methods to extract the local cross-sectional area of the vessel. Using the MISR, which is already provided with the centerline, yields the cross-sectional area, $\text{MISR}^2\cdot\pi$. The exact, but more computationally demanding method, is slicing the geometry at each centerline point and calculating the cross-sectional area by triangulating each slice. Using finite differences, the tangent vector of the centerline path yields the slice normal. Figure~\ref{fig_radius_vs_slice} compares both methods of area extraction for an inlet branch of an aortic geometry. The branch slices are shown on the left, spheres with local MISR are shown in the middle. The graph on the right compares the area calculated from both methods. Here, two deficiencies of the MISR-area are evident. Firstly, the MISR-area underestimates the cross-sectional area at the inlet by a factor of more than two. The MISR is unreliable within the last diameter before the cap surface. Secondly, the MISR-area only offers a lower bound of cross-sectional area in non-circular vessels. This is evident throughout the branch in Figure~\ref{fig_radius_vs_slice}. Only for a perfectly circular vessel, the MISR-area is equal to the cross-section area. We thus use slicing to determine the local cross-sectional area in this study. The slices can also be used to split patient-specific geometries into branches and junctions, as will be shown in Section~\ref{sec_junction}.

\subsubsection{Junction detection \label{sec_junction}}
\begin{figure}[Htp!]
\centering
\includegraphics[width=0.7\textwidth]{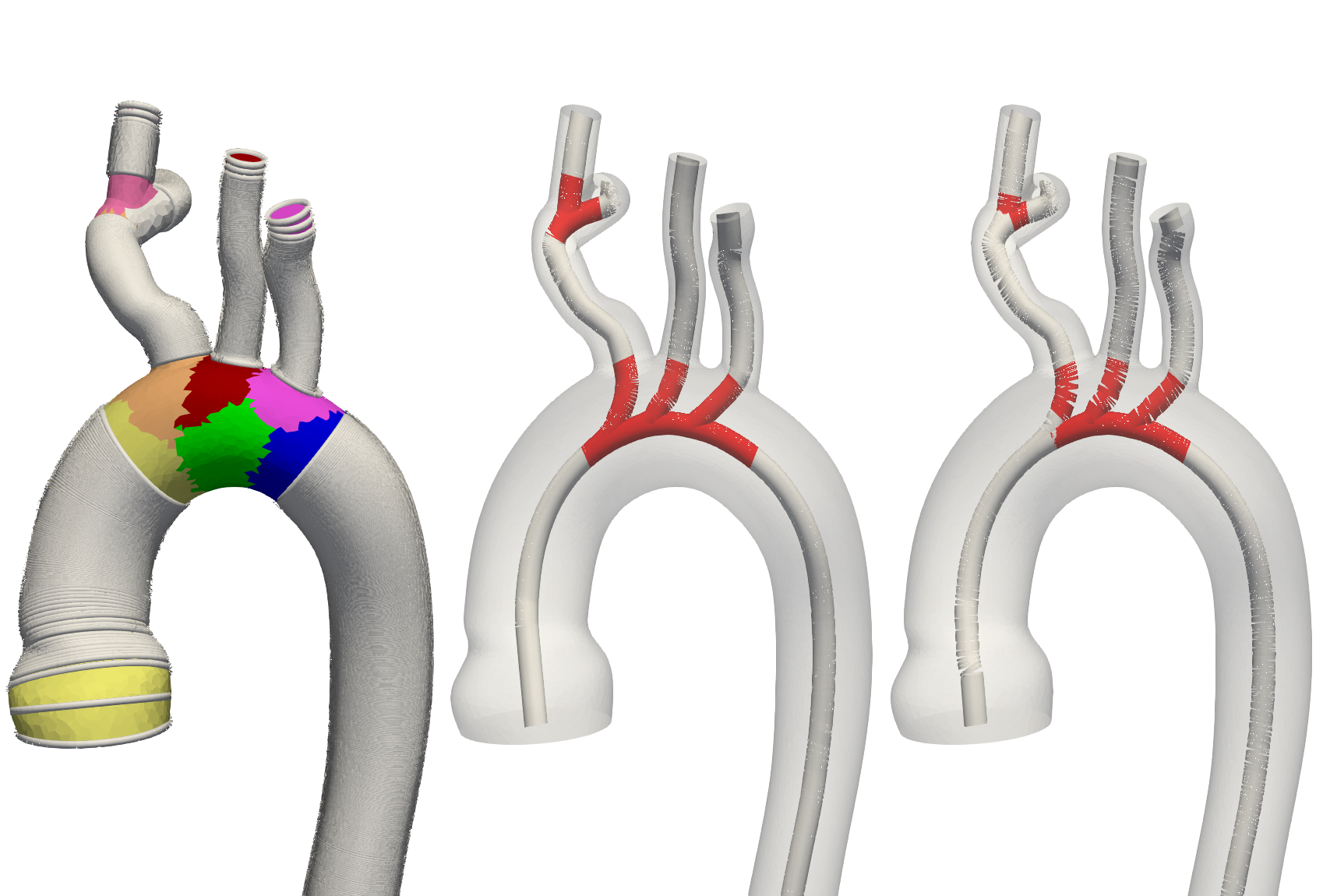}
\caption{The 3D surface of model 0003\_0001 (normal aortafemoral) is colored by proximity to centerline branch segments (left), slices labeled as branches are shown in white. The corresponding split of the centerline (middle) shows branches in white and junctions in red. The junction domains are larger than those detected by VMTK (right). \label{fig_junction}}
\end{figure}

\begin{figure}[Htp!]
\centering
\includegraphics[width=0.7\textwidth]{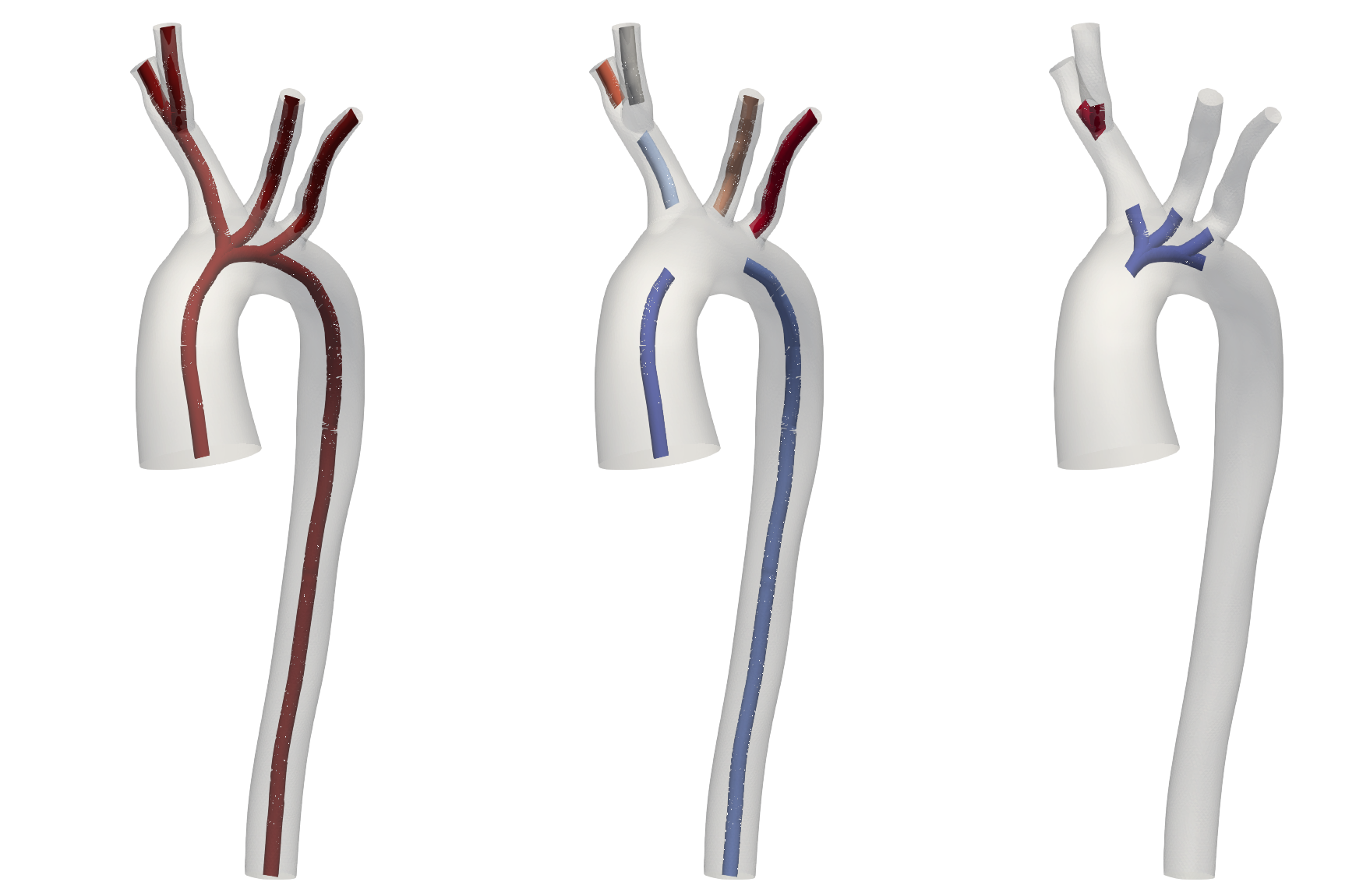}
\caption{Centerline, branches, and junctions (from left to right) in model 0075\_0001 (tricuspid atresia). \label{fig_ids}}
\end{figure}

The theory of 1D and 0D reduced-dimensional blood flow, flow introduced in Sections~\ref{sec_theory_1d} and \ref{sec_theory_0d}, respectively, only applies to individual branch segments. We thus need to split our centerline network into branches and junctions. For that, we employ again the slices of the 3D geometry as explained in Section~\ref{seg_area}. This process is visualized in Figure~\ref{fig_junction}. We color \revb{each point on} the 3D surface based on proximity to centerline branches. We then label all slices that cut through more than one color on the surface as junctions. We also label slices that have more than one centerline passing through them as belonging to a junction. The remaining slices (shown in Figure~\ref{fig_junction}, left) are then labeled as branches. Figure~\ref{fig_junction} (middle) shows the resulting split of the centerline in branches (white) and junctions (red). Note that our junction detection is more comprehensive than the one obtained from VMTK branch extraction (right). Our junction regions are always connected and contain one inlet and at least two outlets. All branches and junctions are finally each assigned a unique identification number, see Figure~\ref{fig_ids}. \revb{Note that this junction criterion depends on surface mesh resolution. A coarser surface mesh will result in larger junction regions and can join  junctions close to each other, e.g., in pulmonary models, to a larger multiple-outlet junction. In this work, we use the surface meshes of published geometries from the VMR models with a mesh resolution chosen for a high quality finite element fluid solution.} \reva{Also note that our junction detection can handle junctions with an arbitrary number of inlets and outlets, as visualized in the aortic arch in Figures~\ref{fig_junction} and \ref{fig_ids}.}

\begin{figure}[Htp!]
\centering
\includegraphics[height=6cm]{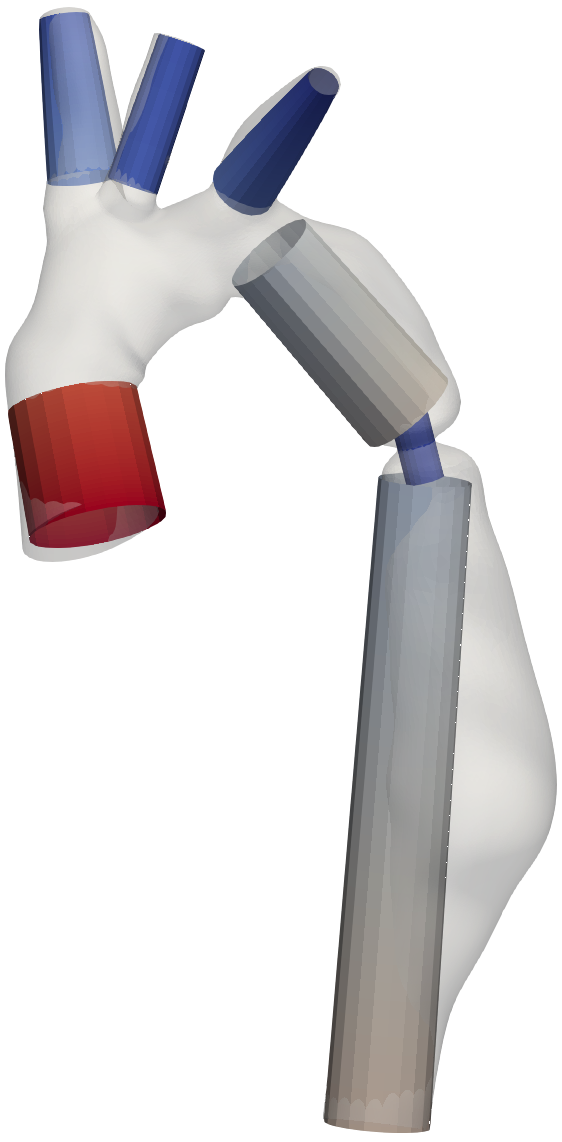}~
\includegraphics[height=6cm]{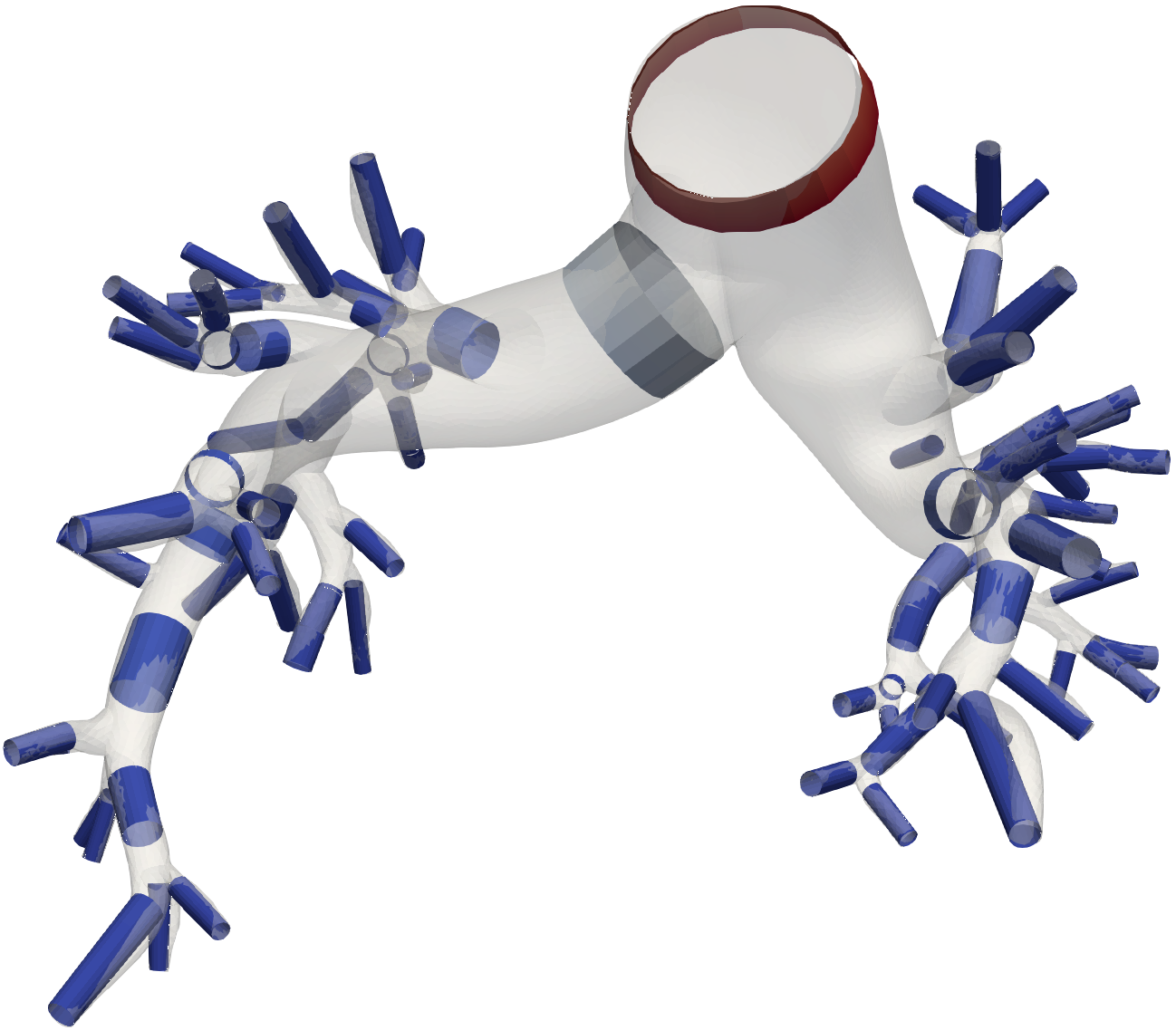}~
\includegraphics[height=6cm]{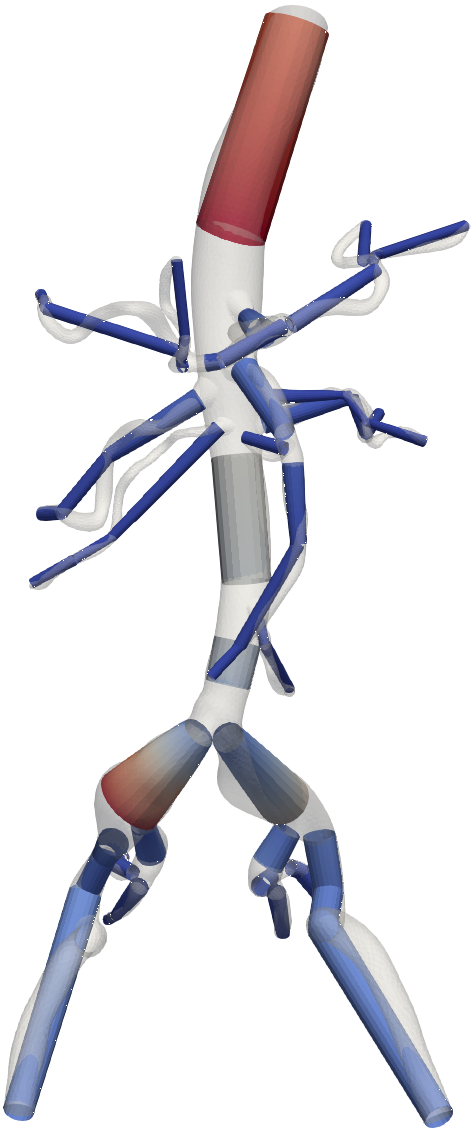}~
\includegraphics[height=6cm]{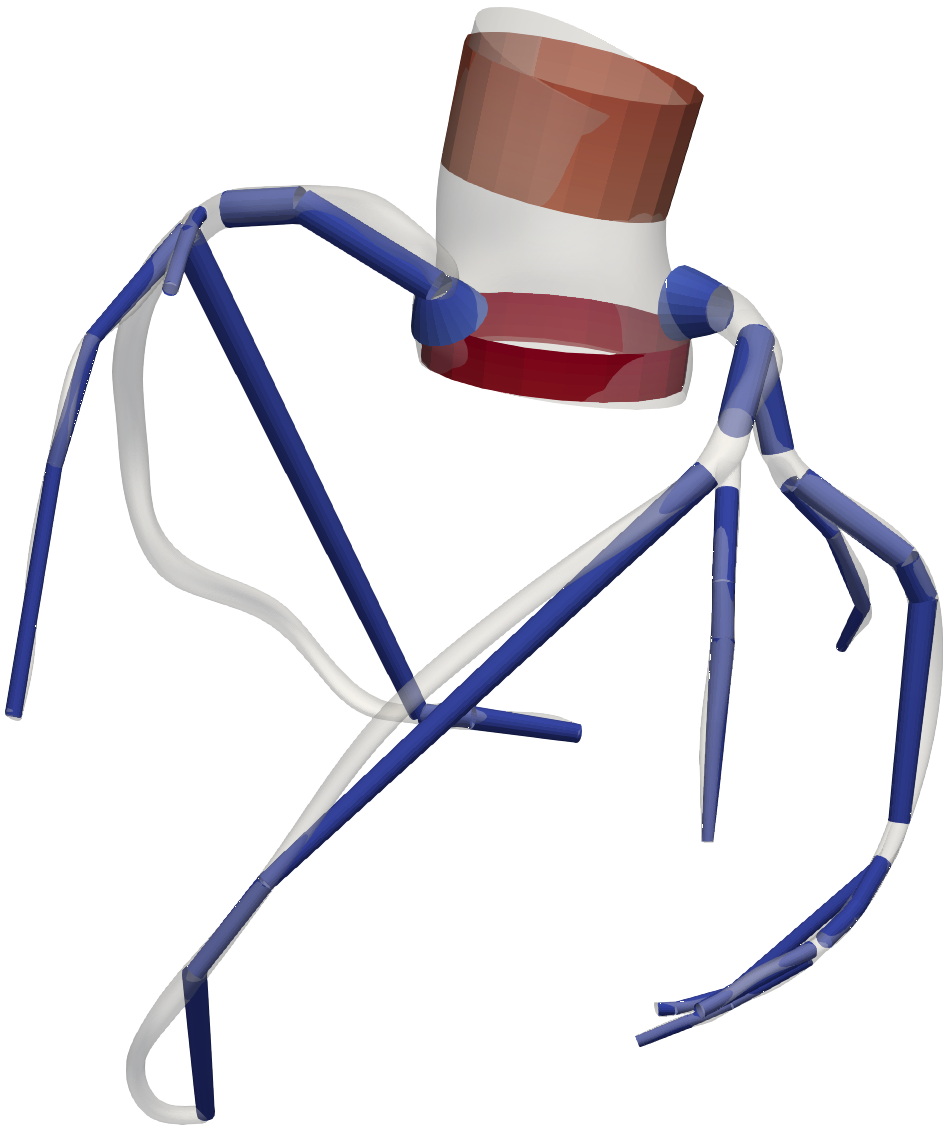}
\caption{\reva{1D models individually colored by cross-sectional area with overlayed 3D surfaces. From left to right models 0069\_0001 (artificial aortic coarctation), 0088\_0001 (pulmonary artery hypertension), 0110\_0001 (normal aortofemoral), 0186\_0002 (normal coronary).}\label{fig_1d_models}}
\end{figure}

\reva{\subsubsection{Stenosis detection \label{sec_stenosis}}}
\reva{Our model generation pipeline offers two modes of 1D and 0D branch discretization: an automatic stenosis-detection offering accurate modeling with minimal computational effort and a user-defined number $n$ of segments per vessel branch. For both 1D and 0D stenosis modeling in \eqref{eqn_1d_stenosis} and \eqref{eqn_0d_stenosis}, respectively, we need to extract stenosed area $S_s$ and proximal cross-sectional area $S_0$. In the automatic mode, we sample the cross-sectional along each branch and extract relative minima and maxima using the function \texttt{argrelextrema} from the \texttt{SciPy} signal processing toolbox. We extract the stenosed area $S_s$ as the relative minima and the proximal area $S_0$ as the corresponding relative maxima. The stenosis length $L$ is determined from the distance to adjacent extrema. The stenosis detection is evident in Figure~\ref{fig_1d_models} where 1D segments are visualized for various anatomies. The stenosis detection is highly visible in the descending aorta of model 0069\_0001 (left, artificial aortic coarctation). To locate the stenosis at the correct location within a vessel branch, we split each vessel into three segments, proximal, stenosis, and distal. In case of several stenoses, for simplicity, we select the one with the largest ratio $S_0/S_s$. In the user-defined mode, we sample the cross-sectional branch area with $n$ segments. We determine their location such that they optimally represent the variation in cross-sectional area, using the Python module \texttt{pwlf}. Here, the ratio $S_0/S_s$ is calculated from the cross-sectional area of adjacent segments. For both modes, the discretization of vessel segments is identical. In 1D, we discretize each segment with a linear interpolation of the cross-sectional area between both branch ends. We create 100 finite elements for each centimeter of vessel length and at least 5 elements per segment. In 0D, we create one 0D element per segment with the vessel radius averaged between both branch ends.}

\section{Results}
In this section, we show the results of our comprehensive comparison of high-fidelity 3D models to 1D and 0D reduced-dimensional models. \reva{Unless stated otherwise, all models are rigid-wall and with automatic stenoses detection in 1D 	and 0D models. To mimic rigid-wall behavior of the 3D models in 1D, we use the Olufsen material model with parameters $k_1=k_3=1000\,\text{kPa}$ and $k_2=-20/\text{cm}$.} \revb{Using these parameters, we verified that the maximum area change in all geometries is below 2\% and we do not encouter problems with the numerical solution of the 1D model.} \reva{For 0D models, we use $RCL$-stenosis elements in each segment with vessel stiffness $k_0=100\,\text{kPa}$.} For comparison, we integrate the 3D solutions over the cross-sections of the vessel. We follow the same approach as in Section~\ref{seg_area} where we extract the cross-sectional area on each centerline point. We then split the vessel into branches and junctions as in Section~\ref{sec_junction}. This allows us to compare the results at the caps of the model, as well as in the interiors, branch by branch.

\subsection{Vascular Model Repository}
\begin{figure}[Htp!]
\centering
\includegraphics[width=.9\textwidth]{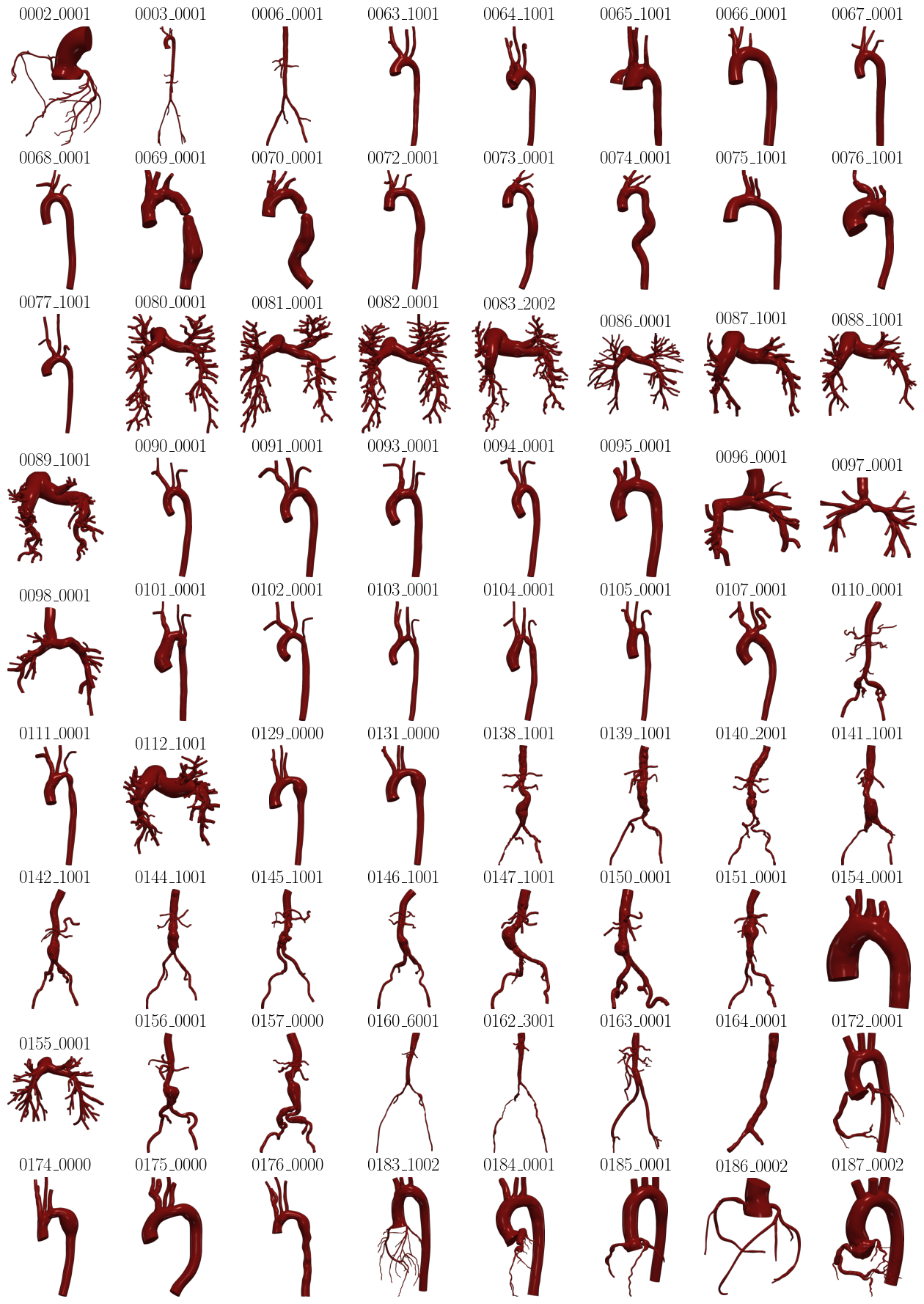}
\caption{Overview of all $\revb{N=72}$ models used in this study. \label{fig_repo_models}}
\end{figure}

\begin{figure}[Htp!]
\centering
\includegraphics[width=\textwidth]{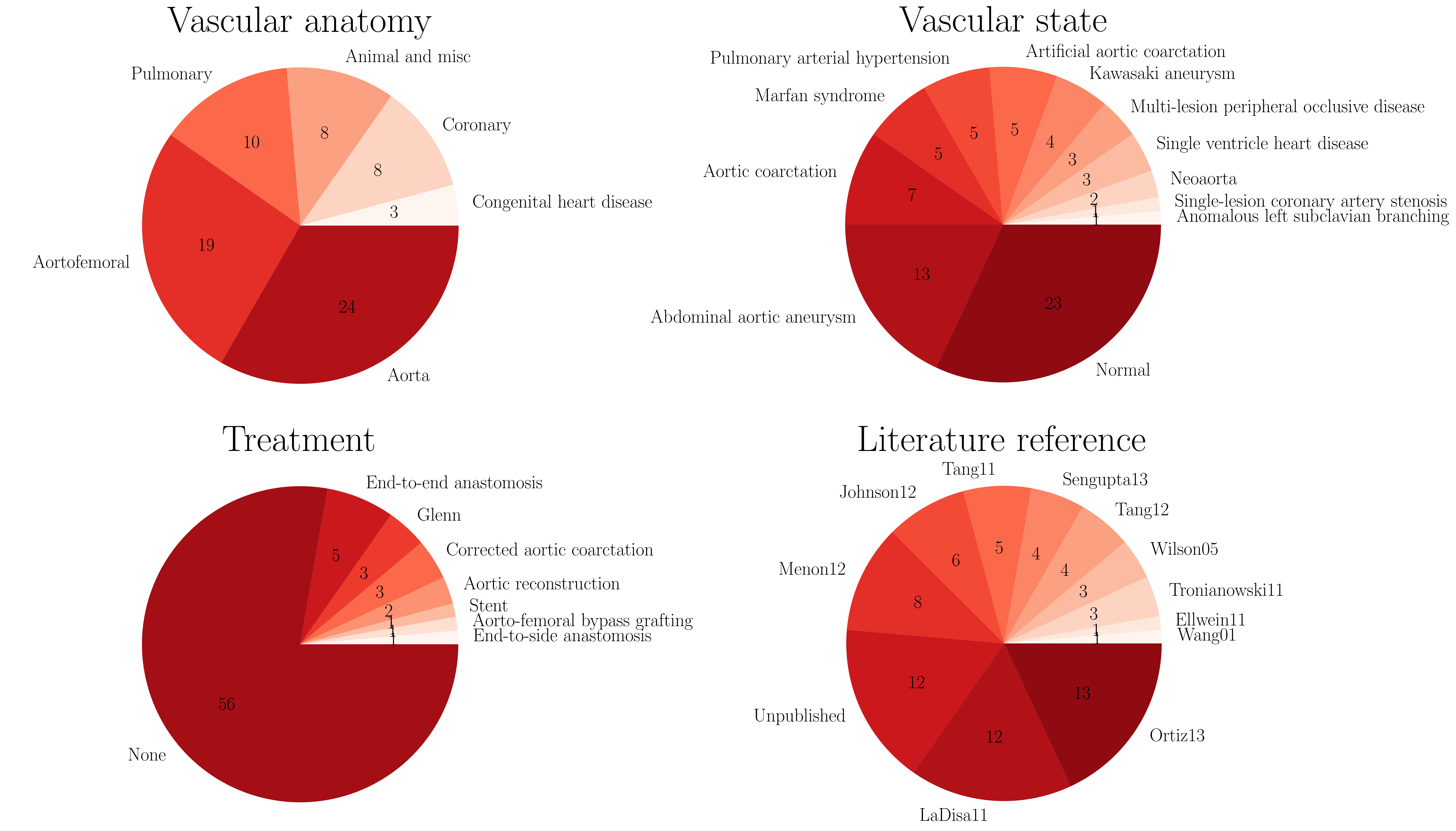}
\caption{Properties of all $\revb{N=72}$ models used in this study. \label{fig_repo_statistics}}
\end{figure}

An overview of all $\revb{N=72}$ geometries used in this work is shown in Figure~\ref{fig_repo_models}. \revb{All our models are freely available at the public Vascular Model Repository (VMR) (see \url{http://vascularmodel.org}) in the form of curated \texttt{SimVascular} projects.\cite{wilson13}}  Figure~\ref{fig_repo_statistics} provides an overview of the model properties. The different anatomies include cerebrovascular,\cite{bockman12} pulmonary arteries in Fontan patients,\cite{marsden10} left circumflex coronary artery,\cite{ellwein11} aortic coarctation (untreated, end-to-end anastomosis),\cite{ladisa11a, ladisa11b, menon11} abdominal aortic aneurysms, coronary artery aneurysms in Kawasaki patients,\cite{sengupta11} superior vena cava and pulmonary arteries in Glenn patients,\cite{troianowski11}  aortoiliac occlusive disease,\cite{wilson05} and several previously unpublished models from these categories. All outlet boundary conditions applied in this work are open-loop LPNs. A pulsatile inflow with a parabolic velocity field is prescribed at each model's inlet cap. The inlet and outlet boundary conditions have been tuned to \textit{in vivo} measurements, such as phase-contrast magnetic resonance imaging and catheter pressure measurements. The majority of the models have three-element Windkessel boundary conditions (RCR).\cite{vignonclementel10} The pulmonary models here generally have resistance boundary conditions with a constant distal pressure. The coronary models use an RCRCR coronary boundary condition with prescribed variable intramyocardial pressures at the outlets.\cite{kim10a, kim10b} The full descriptions of all models and the outlet boundary conditions are provided on the VMR website. 

\subsection{Computational performance}
Using the high-performance computing setup detailed in Section~\ref{sec_theory_3d}, simulating a single 3D model requires about two days of computation time. All ROM generation and computing was performed on a single CPU on a workstation computer. Generating the centerlines for all $\revb{N=72}$ models took  \revb{3.6\,h (mean 3.0\,min, std. 2.9\,min)}. The centerlines need to be extracted only once per geometry, the rest of the 1D and 0D model generation pipeline is instantaneous. We estimated the number of cardiac cycles required to reach a periodic state from initial conditions generated from a mean flow solution based on the results in a prior study.\cite{pfaller21} \revb{The runtimes of the models are summarized in Table~\ref{tab_errors}. On average, 0D models took 0.84 minutes to compute and 1D models about six times as long.}

\begin{table}[Htp!]
\begin{center}
\begin{tabular}{ |c|c|c|c|c|c|c|c|c|c| }
\hline
 & \multicolumn{2}{c|}{\textbf{average error [\%]}} & \multicolumn{2}{c|}{\textbf{maximum error [\%]}} & \multicolumn{2}{c|}{\textbf{systolic error [\%]}} & \multicolumn{2}{c|}{\textbf{diastolic error [\%]}} & \textbf{runtime [min]}\\& pressure& flow& pressure& flow& pressure& flow& pressure& flow & \\
\hline
\textbf{1D} & 1.8 / 1.7 & 3.4 / 2.7 & 5.9 / 6.2 & 11 / 9.9 & 4.2 / 5.1 & 7.9 / 7.7 & 1.1 / 1.1 & 2.9 / 2.4 & 4.9 / 12 \\
\hline
\textbf{0D} & 2.1 / 1.6 & 3.9 / 2.7 & 6.9 / 5.2 & 13 / 7.4 & 5.2 / 4.5 & 9.4 / 6.8 & 1.2 / 1.2 & 3.7 / 3.5 & 0.84 / 1.2 \\
\hline
\end{tabular}
\end{center}
\caption{\reva{Average of all $N=72$ relative errors and runtimes at caps in 1D and 0D models (mean / standard deviation). \label{tab_errors}}}
\end{table}

\subsection{Approximation quality \label{sec_quality}}
Pressure, $P_{t,i}^\text{3D}$, and flow, $Q_{t,i}^\text{3D}$, are extracted at time step $t$ and centerline branch point $i$. The 3D solution provides the ground-truth reference solution for both 1D and 0D simulations.  We compare 3D to $d\in\{1,0\}$ models using the relative \reva{cap} error metrics \reva{averaged and maximum over all time steps and during systole and diastole}:\cite{alastruey11,xiao13}
\begin{alignat}{4}
\epsilon^{d\text{D}}_{P,\text{avg}} &= \frac{1}{n_\text{cap}} \sum_{i=1}^{n_\text{cap}} \frac{\sum_{t=1}^{n_t} \left| P^{d\text{D}}_{t,i} - P_{t,i}^{3\text{D}}\right|}{\sum_{t=1}^{n_t} P_{t,i}^\text{3D}}, \quad &&
\epsilon^{d\text{D}}_{Q,\text{avg}} &&= \frac{1}{n_\text{cap}n_t} &&\sum_{i=1}^{n_\text{cap}} \frac{\sum_{t=1}^{n_t} \left| Q^{d\text{D}}_{t,i} - Q_{t,i}^{3\text{D}}\right|}{\max_t Q_{t,i}^\text{3D} - \min_t Q_{t,i}^\text{3D}}, \\
\epsilon^{d\text{D}}_{P,\text{max}} &= \frac{n_t}{n_\text{cap}} \sum_{i=1}^{n_\text{cap}} \frac{\max_t \left| P^{d\text{D}}_{t,i} - P_{t,i}^{3\text{D}}\right|}{\sum_{t=1}^{n_t} P_{t,i}^\text{3D}}, \quad &&
\epsilon^{d\text{D}}_{Q,\text{max}} &&= \frac{1}{n_\text{cap}} &&\sum_{i=1}^{n_\text{cap}} \frac{\max_t \left| Q^{d\text{D}}_{t,i} - Q_{t,i}^{3\text{D}}\right|}{\max_t Q_{t,i}^\text{3D} - \min_t Q_{t,i}^\text{3D}},\\
\epsilon^{d\text{D}}_{P,\text{sys}} &= \frac{n_t}{n_\text{cap}} \sum_{i=1}^{n_\text{cap}} \frac{\left| P^{d\text{D}}_{t_\text{sys},i} - P_{t_\text{sys},i}^{3\text{D}}\right|}{\sum_{t=1}^{n_t} P_{t,i}^\text{3D}}, \quad &&
\epsilon^{d\text{D}}_{Q,\text{sys}} &&= \frac{1}{n_\text{cap}} &&\sum_{i=1}^{n_\text{cap}} \frac{\left| Q^{d\text{D}}_{t_\text{sys},i} - Q_{t_\text{sys},i}^{3\text{D}}\right|}{\max_t Q_{t,i}^\text{3D} - \min_t Q_{t,i}^\text{3D}},
\quad t_\text{sys} = \text{arg\,max}_t \, Q_{t,\text{inlet}}^{3\text{D}}\\
\epsilon^{d\text{D}}_{P,\text{dia}} &= \frac{n_t}{n_\text{cap}} \sum_{i=1}^{n_\text{cap}} \frac{\left| P^{d\text{D}}_{t_\text{dia},i} - P_{t_\text{dia},i}^{3\text{D}}\right|}{\sum_{t=1}^{n_t} P_{t,i}^\text{3D}}, \quad &&
\epsilon^{d\text{D}}_{Q,\text{dia}} &&= \frac{1}{n_\text{cap}} &&\sum_{i=1}^{n_\text{cap}} \frac{\left| Q^{d\text{D}}_{t_\text{dia},i} - Q_{t_\text{dia},i}^{3\text{D}}\right|}{\max_t Q_{t,i}^\text{3D} - \min_t Q_{t,i}^\text{3D}},
\quad t_\text{dia} = \text{arg\,min}_t \, Q_{t,\text{inlet}}^{3\text{D}}
\end{alignat}
with number of time steps $n_t$ and number of \reva{cap points $n_\text{cap}$}. Note that we exclude the inflow $Q_{t,\text{inlet}}^{3\text{D}}$ when calculating flow errors since it is prescribed. The pressure difference is normalized at each cap by the 3D time-averaged pressure whereas the flow difference is normalized by the flow amplitude. \revb{For plotting over vessel branches,} we \revb{branch-wise} linearly interpolate 1D and 0D pressure and flow, $P_{t,i}^\text{dD}$ and $Q_{t,i}^\text{dD}$, respectively, onto the centerline points \revb{using the centerline branch path length}. \revb{This interpolation is necessary since 1D and 0D results are only available at finite element nodes and at branch ends, respectively.}

\begin{figure}[Htp!]
\centering
\includegraphics[width=.5\textwidth]{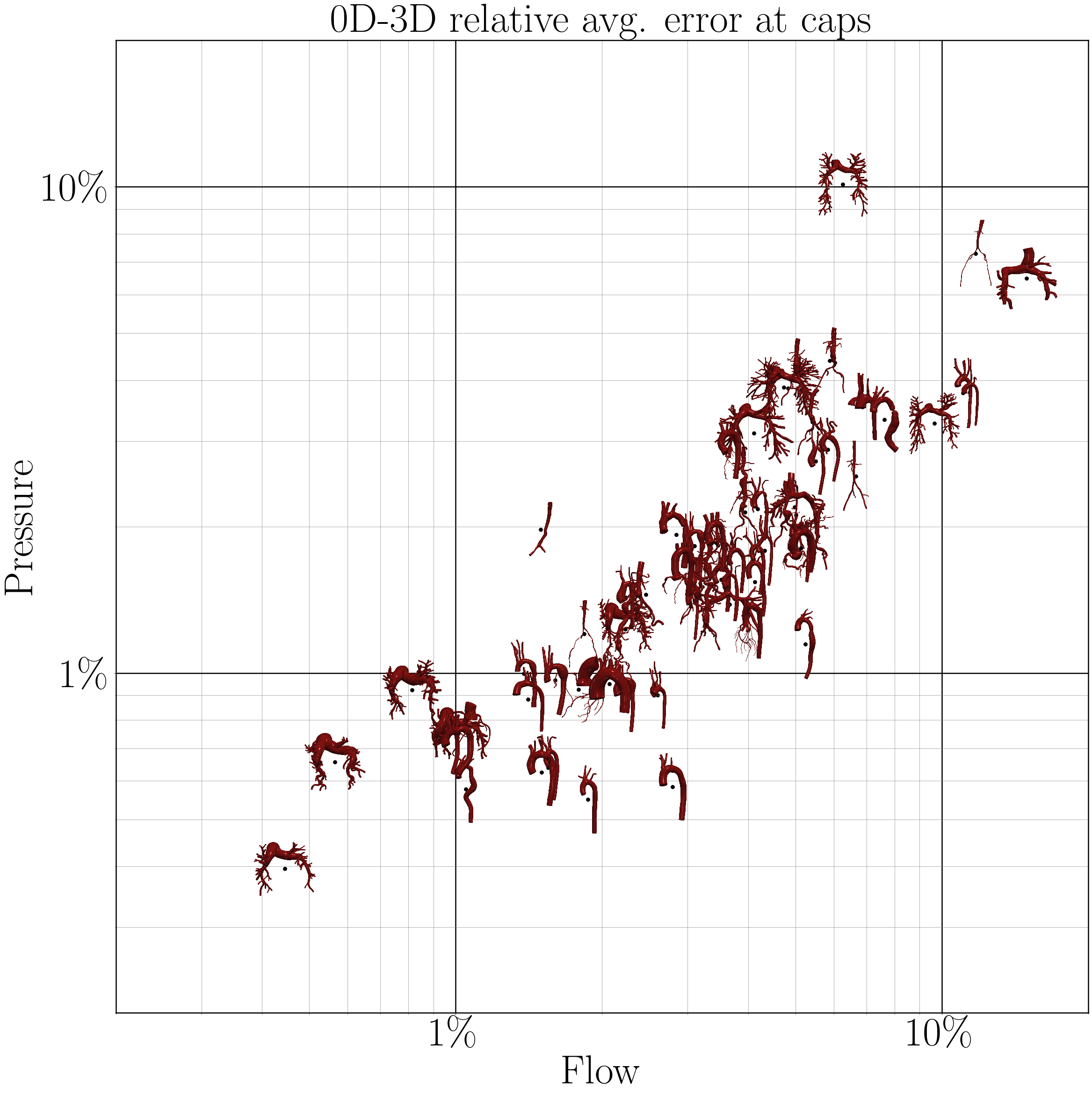}~
\includegraphics[width=.5\textwidth]{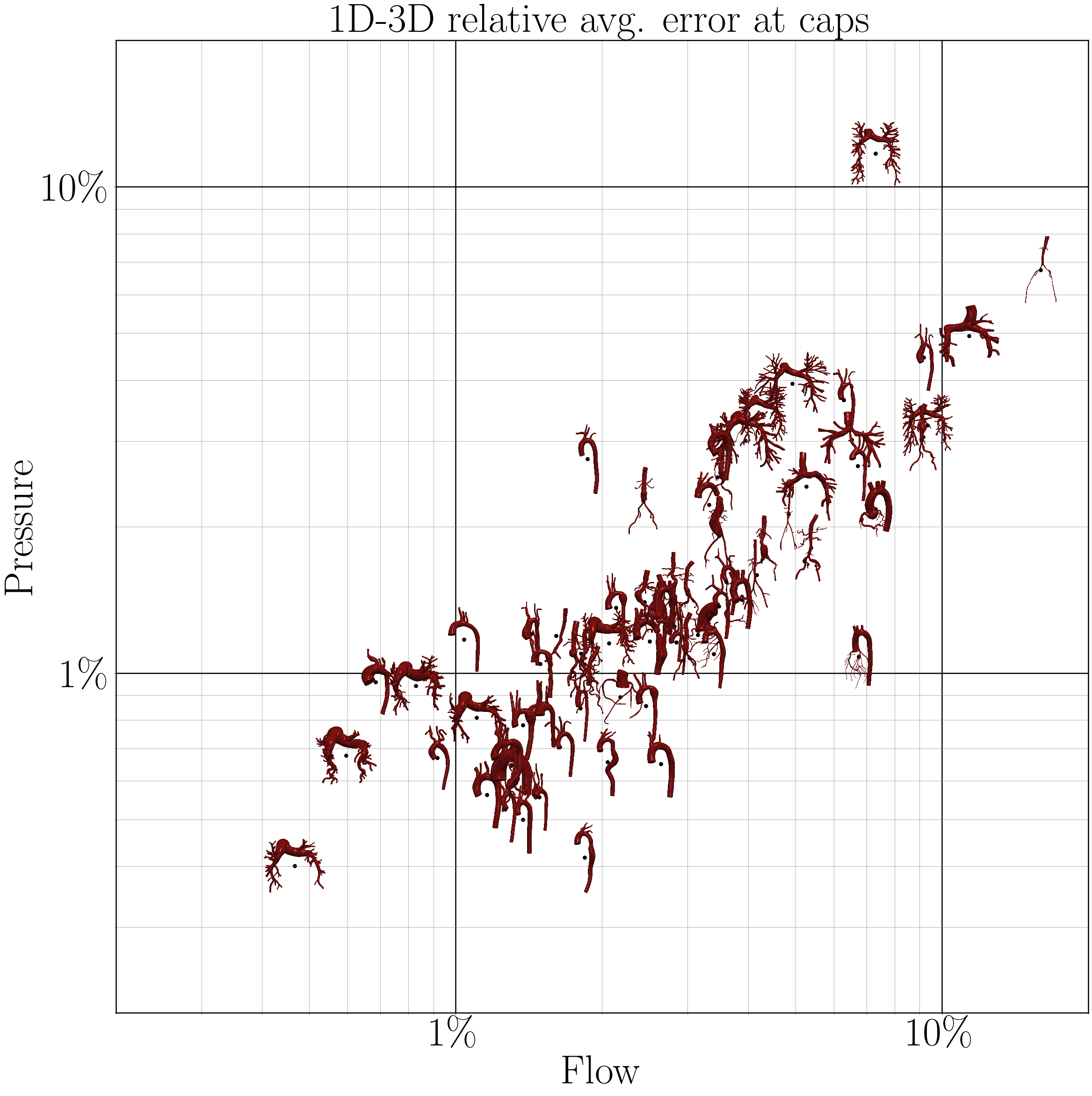}\\
\includegraphics[width=.5\textwidth]{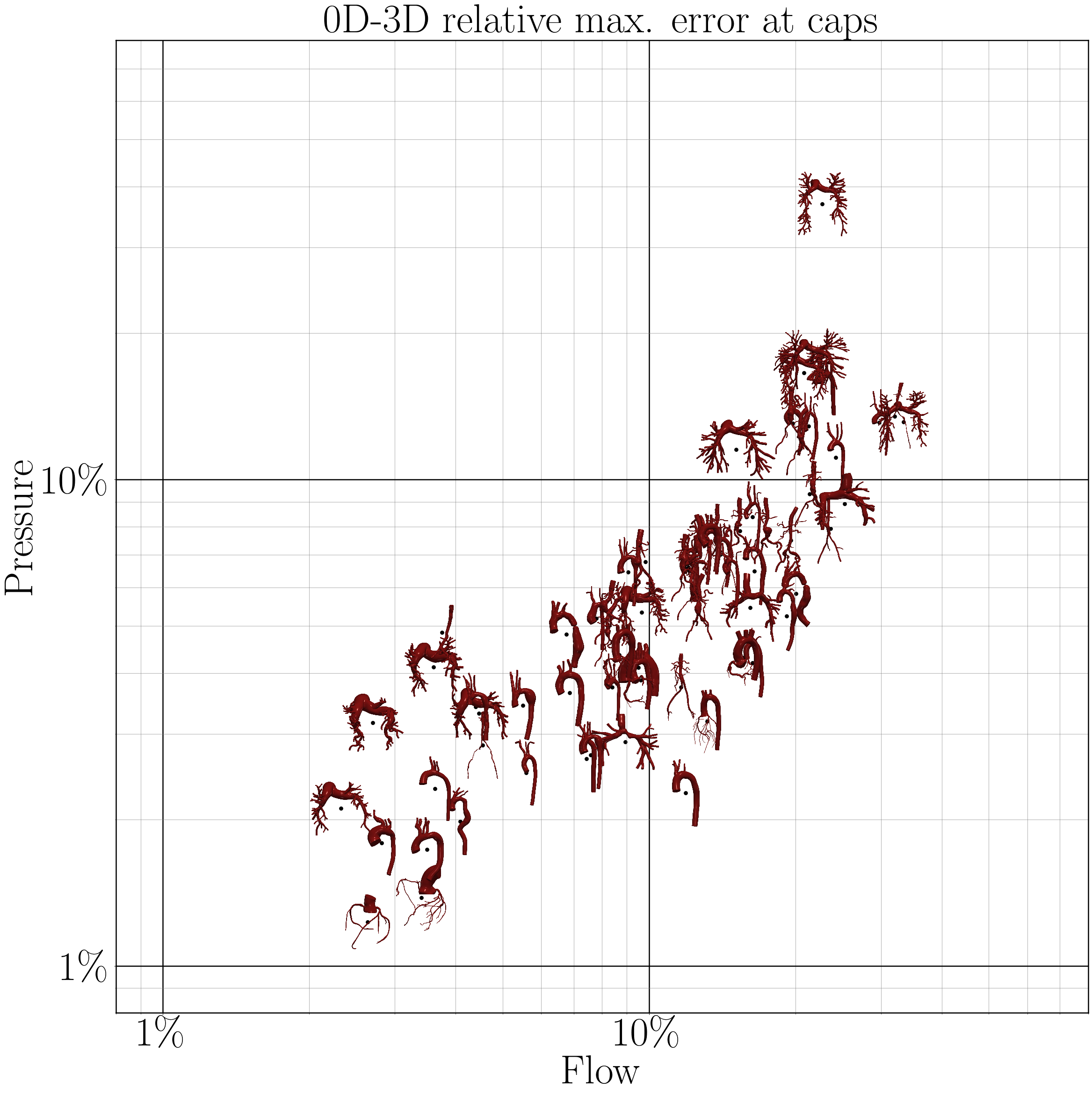}~
\includegraphics[width=.5\textwidth]{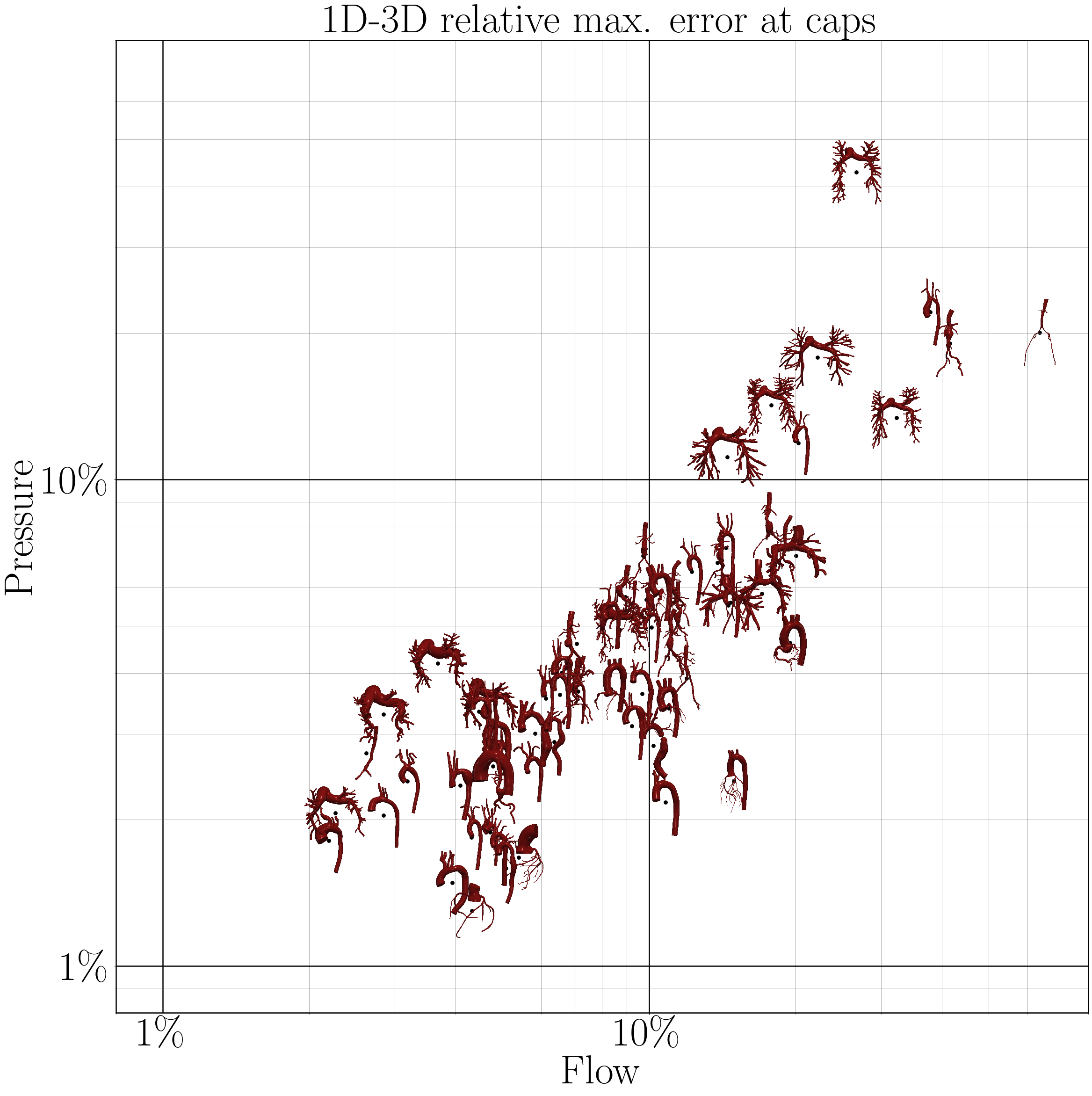}
\caption{\reva{Correlation between average (top) and maximum (bottom) flow and pressure error at the caps of the model in 0D (left) and 1D simulations (right) compared to 3D simulations. \label{fig_error_correlation_flow_pressure_avg_rel_img}}}
\end{figure}

The flow and pressure errors at the caps for all $\revb{N=72}$ models are shown in Figure~\ref{fig_error_correlation_flow_pressure_avg_rel_img}. Models appearing in the lower left of each plot are approximated well, while models in the top right have higher errors. A majority of the models falls in the range $1\%<\epsilon<10\%$ with some exceptions above and below. Examining results of 0D models, some pulmonary models fall outside this range. This is due to the fact that the models with low approximation error are patients with pulmonary arterial hypertension. In those models, the flow is dominated by the resistances and distal pressures of the boundary conditions, more than the 3D fluid dynamics inside the model. These models are thus well represented as ROMs. Normal pulmonary models exhibit high approximation errors since these models consist of a large number of subsequent vessel junctions, which are currently not modeled in our 1D or 0D models. \revb{Some models with severe stenoses within junction exhibit errors close to $10\%$ since stenoses within junctions are currently not included in our modeling framework.}

\reva{The averaged errors for all $N=72$ models in this study are summarized in Table~\ref{tab_errors}. In general, 0D errors are only slightly higher than 1D errors despite taking only a third of the 1D runtime. Pressure is approximated better than flow, although this also depends on the choice of error normalization (average vs. amplitude). Diastolic quantities are approximated much better than systolic quantities. This is expected since flow is minimal during diastole. Thus, fluid dynamics are mainly dictated by boundary conditions, which are identical in all model fidelities, rather than the fluid models themselves. Furthermore, flow and pressure are generally lowest in diastole, resulting in a low relative error.}

\begin{figure}[Htp!]
\centering
\includegraphics[width=\textwidth]{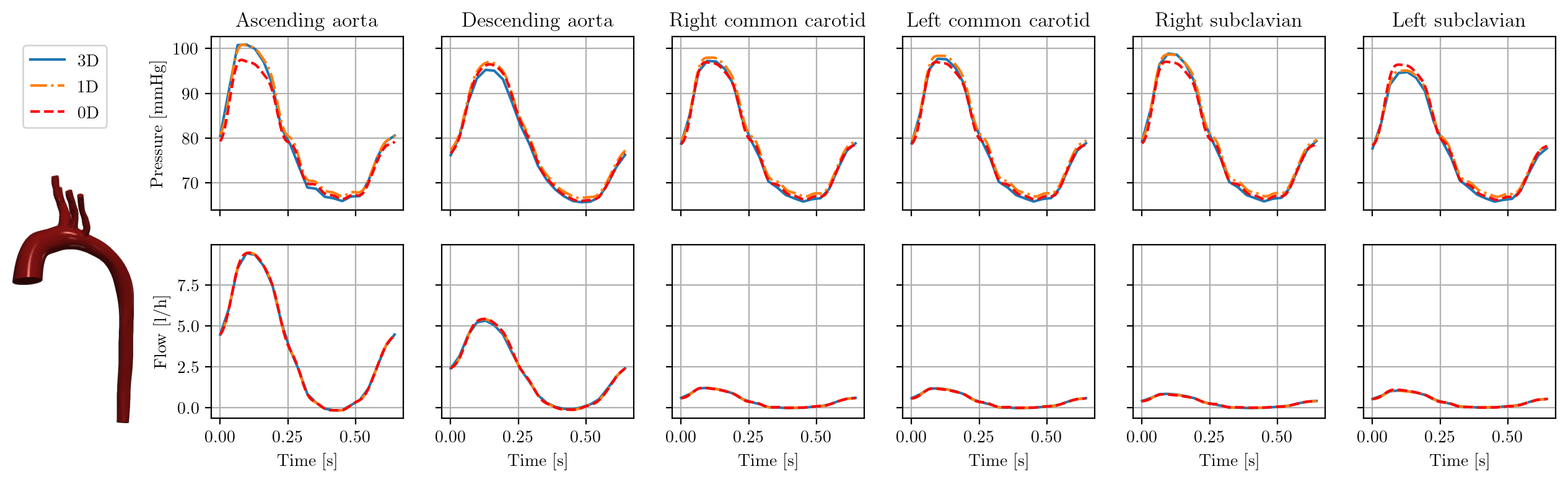}
\caption{\reva{Cap results pressure (top) and flow (bottom) in model 0075\_1001 (tricuspid atresia) over one cardiac cycle.\label{fig_0075_1001_caps}}}
\end{figure}

\begin{figure}[Htp!]
\centering
\includegraphics[width=\textwidth]{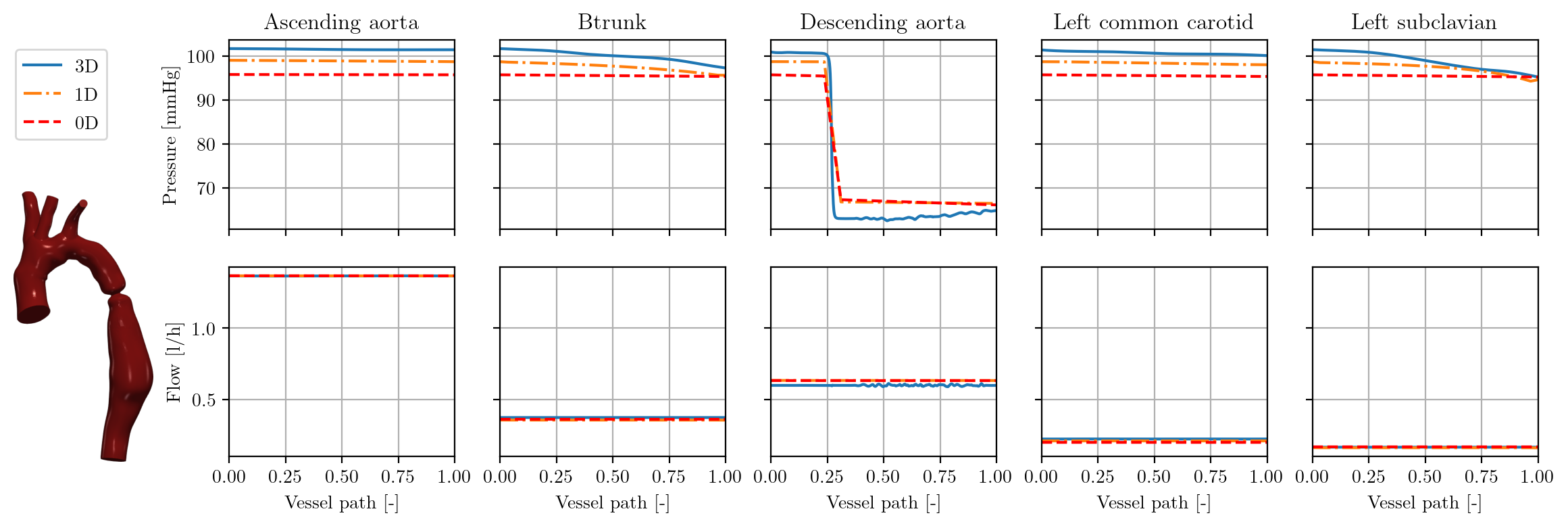}
\caption{\reva{Interior results pressure (top) and flow (bottom) in model 0069\_0001 (severe aortic coarctation in the descending aorta) at peak systole over vessel paths.\label{fig_0069_0001}}}
\end{figure}

\begin{figure}[Htp!]
\centering
\includegraphics[width=\textwidth]{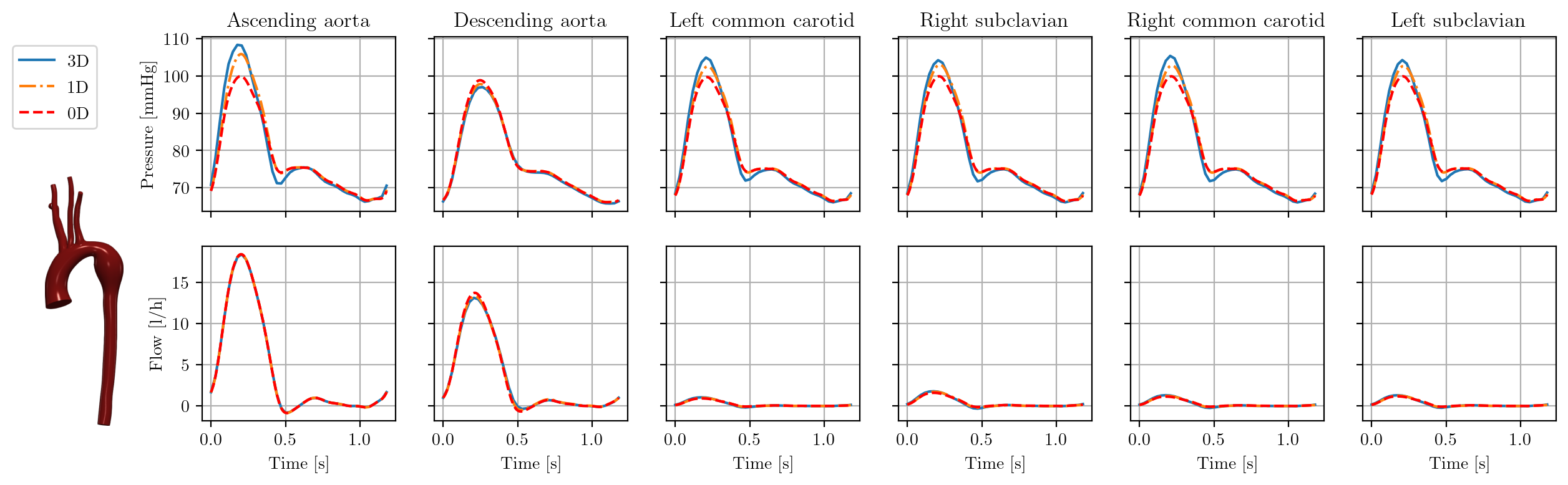}\\
\includegraphics[width=\textwidth]{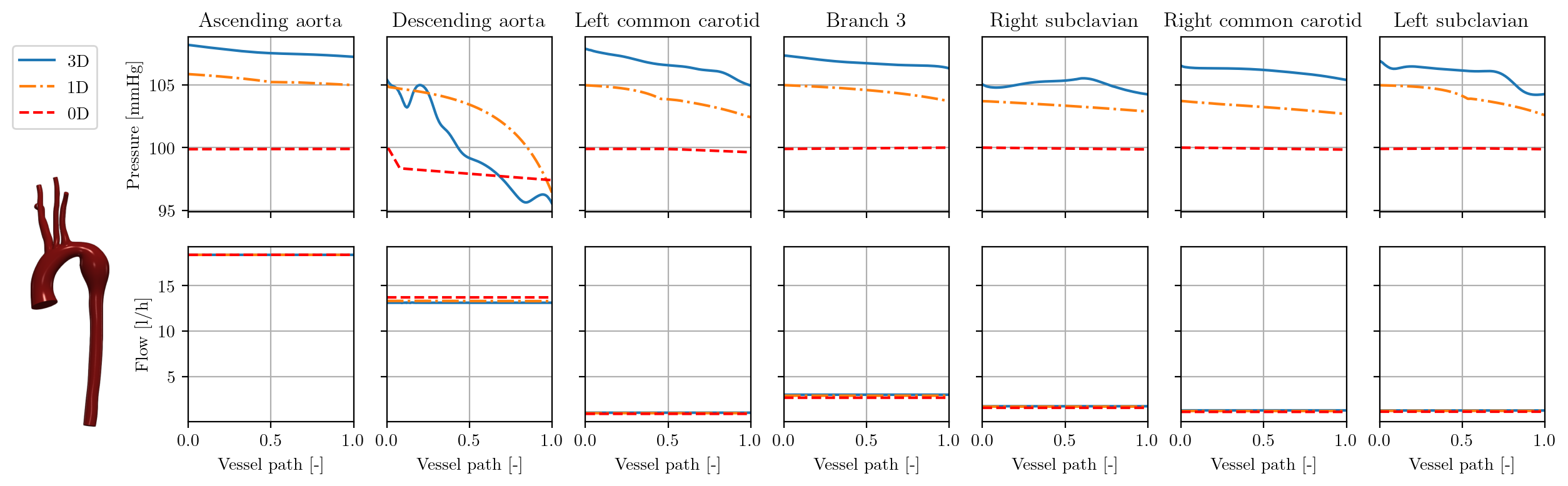}
\caption{\reva{Results in model 0129\_0000 (Marfan syndrome): caps over one cardiac cycle (top), interior at peak systole over vessel paths (bottom).\label{fig_0129_0000}}}
\end{figure}

In Figures \ref{fig_0075_1001_caps} to \ref{fig_0129_0000}, we examine individual cases. Individual results are shown for model 0075\_1001 in Figure~\ref{fig_0075_1001_caps}. In this normal aorta geometry, flow and pressure are approximated well for both 1D and 0D. Pressure curves match the 3D solution well, with 0D slightly underestimating pressure. Flow curves and flow splits between outlets are approximated are almost identical in all three models.

\reva{Figure~\ref{fig_0069_0001} (top) shows model 0069\_0001, a small animal model of an artificially generated severe coarctation in the descending aorta (see Figure~\ref{fig_1d_models}, left). The severe stenosis is evident as a rapid pressure drop in the descending aorta. This pressure drop is well represented by 1D and 0D model. The 3D model predicts a pressure drop over the coarctation of 37\,mmHg. This is estimated as 32\,mmHg and 27\,mmHg in 1D and 0D, respectively, at roughly the correct location along the vessel. Note that without 1D and 0D stenosis modeling detailed in Section~\ref{sec_stenosis}, there would not be a sudden pressure drop in the descending aorta. Instead, there would be a linear pressure drop over the whole length of the descending aorta of about 2\,mmHg. Thus, results would highly overestimate and underestimate pressures distal and proximal to the stenosis, respectively.}

Figure~\ref{fig_0129_0000} (top) shows the cap results for an aortic model 0129\_0000 with a dilated descending aorta in a patient with Marfan syndrome. Similar to the case of aortic coarctation, 1D and 0D models slightly underestimate the pressure in the 3D model. Observing the results in the interior of the model in Figure~\ref{fig_0129_0000} (bottom), it becomes evident that \reva{only the 1D model} accurately represent the 3D pressure drop of $10$\,mmHg. Here, the ROMs cannot fully capture detailed 3D flow features like recirculation in the aneurysm.   

\subsection{\reva{Branch refinement}}
\begin{figure}[Htp!]
\centering
\includegraphics[width=\textwidth]{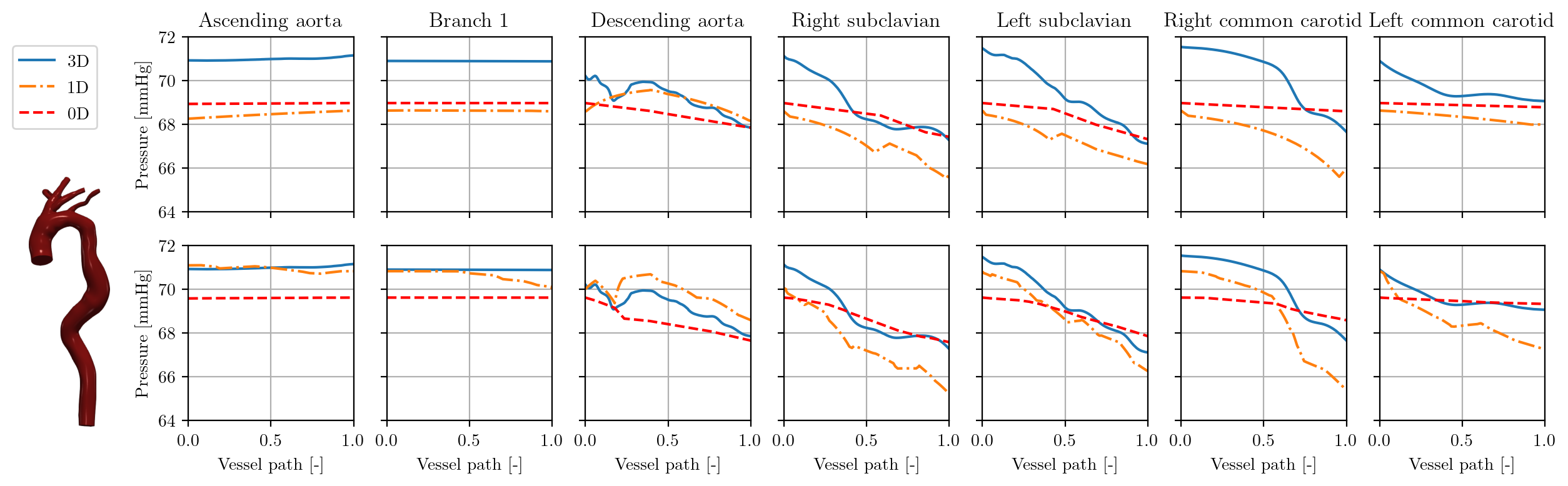}
\caption{\reva{Results in model 0074\_0001 (corrected artificial aortic coarctation): interior pressure at peak systole over vessel paths with automated stenosis detection (top) and $n=10$ segments per vessel branch (bottom).\label{fig_0074_0001_refinement}}}
\end{figure}

\reva{While all results in Section~\ref{sec_quality} where computed with automatic stenosis detection, we compare in Figure~\ref{fig_0074_0001_refinement} automatic results (top) to a user-defined segmentation with $n=10$ segments per vessel branch (bottom) for model 0074\_0001 (corrected artificial aortic coarctation). In the refined case, both 1D and 0D solution approximate 3D pressures better in all branches.  The differences in pressure between both discretizations are most pronounced in the descending aorta, where there is a lot of variation of cross-sectional area over the length of the vessel branch. Here, the 1D solution approximates the local pressure variations along the vessel path better than with automatic stenoses detection. However, this local pressure variation cannot be represented by the refined 0D solution. The runtimes in the refined case are a factor of 11 and 16 longer in 1D and 0D simulations, respectively.}

\subsection{\reva{Deformable walls \label{sec_fsi}}}
\begin{figure}[Htp!]
\centering
\includegraphics[width=\textwidth]{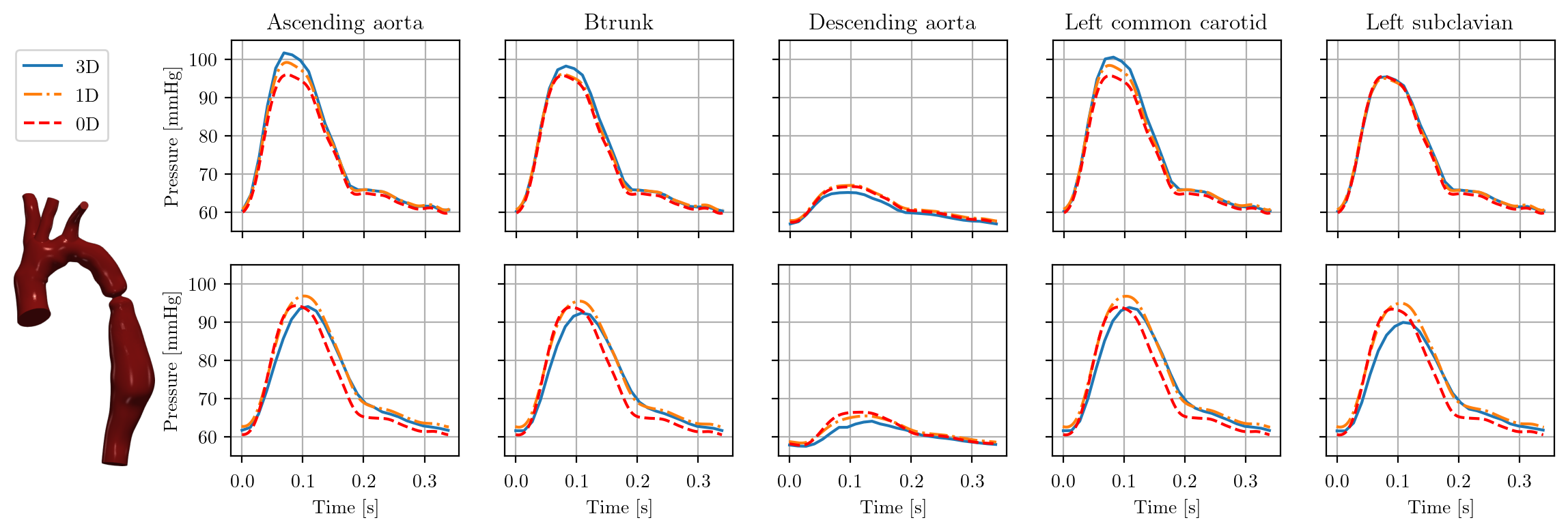}
\caption{\reva{Cap pressure in model 0069\_0001 (severe aortic coarctation) over one cardiac cycle with rigid (top) and compliant walls (bottom).\label{fig_0069_0001_deformable}}}
\end{figure}

\reva{Given the rigid-wall nature of all models in the VMR (at time of publication), we show an additional example of our reduced-order pipeline using a deformable-wall model. Here, we simulate model 0069\_0001 (severe aortic coarctation) with deformable walls and identical geometry and boundary conditions. In 3D, we utilize the coupled momentum method,\cite{figueroa06} which is a linearized approach to fluid-structure interaction using a fixed mesh. We assume the local wall thickness $h$ to be 10\% of the radius $r$ and Young's modulus to be $300 \,\text{kPa}$. We thus obtain a constant ratio $Eh/r=30\,\text{kPa}$ which is equal to the material constant $k_0$ in the 1D linear material model in \eqref{eqn_1d_mat} and is used to calculate the capacitance $C$ in \eqref{eqn_0d_mat}. Figure~\ref{fig_0069_0001_deformable} shows the pressure over time at all outlets. The maximum change in 1D cross-sectional area is now 30\%. The pressure peaks in all three model fidelities are more pronounced and appear earlier in the cardiac cycle with rigid walls (top) compared to deformable walls (bottom). The approximation quality is similar in both cases, with 1D and 0D underestimating pressure in the rigid case and overestimating it in the deformable case.}

\section{Discussion \label{sec_discussion}}
We presented a robust and fully automated pipeline to generate reduced-order 1D and 0D models from 3D geometries. This framework is openly available and integrated in the \texttt{SimVascular} graphical user interface as well as the Python interface. This allows for free and easy use, even in more advanced applications where 0D and/or 1D ROMs are integrated in a user-defined framework, such as uncertainty quantification and parameter estimation. Additionally, our \texttt{svZeroDSolver} is highly modular, allowing users to easily define custom 0D elements.

In general, despite the significantly reduced cost, bulk flow and pressure waveforms on the caps and inside branches are predicted well by both 1D and 0D models. This holds for a large variety of anatomies, vessel types, and disease conditions. For most of the models included in this study, the average approximation error of 1D and 0D models was mostly below $10\%$ compared to 3D fluid dynamics. In general, 1D models performed slightly better than 0D models. \revb{The 1D formulation takes into account wave propagation which is missing in 0D, although this is expected to play only a minor role here due to the quasi-rigid behavior of the walls.} However, the slightly increased performance comes at \revb{roughly six-fold} computational costs. \revc{With the current 0D Python implementation and the 1D C++ implementation, we expect future performance increased of the 0D solver when it is also implemented in C++.} \revb{In general, we have found that the robustness of our 0D solver to be superior to our 1D solver. While we were able to obtain a solution for all $N=72$ models in this study, we encountered convergence issues in geometries that included many large changes in cross-sectional areas (tapering or expanding) over a vessel branch.}

\revb{The 1D and 0D models take into account pressure losses due to sudden changes in cross-sectional area. We achieved a good approximation of the 3D pressure solutions in this work, even in severely (artificially generated) stenoses. However, the model is highly dependent on the choice of cross-sectional area sampling locations.} In addition to predicting the pressure drop in a stenosed segment, we also locate it along a vessel branch. Note that this does not change the overall stenosis resistance of a branch. The stenosis placement thus does not have any influence on the cap results of the 0D simulation but only in the interior of the model.

As part of our automated framework, we introduced a method to split any vessel centerline into branches and junctions, assigning unique identification numbers to each. This allows for easy ROM generation as a connectivity tree can be directly generated from the centerline. In addition, it simplifies the post-processing of the 3D solutions and allows one to display results along specific branches and compare them to ROMs. Splitting the model into branches and junctions allows us to distinguish between different components of the ROMs. Flow in the branches is then approximated by 1D or 0D models. \reva{We assumed that static pressure is conserved in 1D and 0D junctions, similar to previous studies.\cite{stergiopulos92,olufsen99,wan02,steele03,reymond09} Other models aim to preserve total pressure, i.e. static plus dynamic pressure.\cite{matthys07,mynard15,mueller19} Furthermore, models exist to predict the pressure losses over junctions.\cite{mynard15,chnafa17,blonski20} However, in our preliminary studies, these did not lead to a significant improvement in approximation quality and were thus not included. Furthermore, we found little difference in accuracy over preserving total vs. static pressure. In our experience, regional variations in cross-sectional area and their representation in 1D and 0D models had the strongest influence the accuracy of flow and pressure approximation.} Thus, future work could feature a machine-learning-based junction \revb{or stenosis model \cite{fossan21}}, trained on the \revb{3D solutions} in our database.

\reva{At time of publication,} all 3D models in the VMR are rigid-wall, whereas the 1D formulation is inherently deformable-wall. Our choice of using a deformable-wall formulation with a high wall stiffness inherently yields a worse approximation than a truly rigid formulation. \reva{In one modified model, we demonstrated that our framework also yields good approximations of the 3D solution in a deformable-wall simulation.}

Currently, centerline extraction is the bottleneck of the ROM generation pipeline, specifically the extraction of the cross-sectional area at all centerline points, which can take several minutes for a large geometry. This is not a significant limitation in practice since the centerline needs to be generated only once per model. \revb{Currently, the centerline discretization size is determined by the surface mesh size of the 3D geometry.} Future improvements include reducing the number of centerline points where the cross-section is extracted and speeding up the slicing of the 3D geometry by using isosurfaces. 

\reva{In this study, we chose a minimal discretization size for our 1D and 0D models of a maximum of three segments per vessel branch to locate possible stenoses. In an additional example, we demonstrated the effects of manually refining the number of segments per vessel branch, allowing a finer sampling of the cross-sectional area along the branches. In the case of the 1D simulation, this yielded a slightly more accurate local pressure distribution along vessel branches. However, this came at the cost of a significantly increased computational effort.} Future ROMs will have to combine predicting accurate local pressure variations with computational efficiency. A model based on the Port-Hamiltonian method might offer a distributed lumped-parameter hybrid model between our 1D and 0D models.\cite{mora20} Additionally, instead of physics-based ROMs, flow and pressure along the centerline could also be predicted by a neural network. \revc{For any future ROM developments, this study can serve as a baseline for ROM accuracy. Furthermore, this pipeline could be used to quickly curate or assess quality of 3D simulations in future work. This would be particularly relevant to expansion of the VMR with datasets from outside groups.} We also plan to define a coupling interface between 3D solvers and \texttt{svZeroDSolver}, creating a single modular framework to define complex open-loop and closed-loop boundary conditions.

\section*{Acknowledgments \label{sec_acknowledgements}}
We thank Dr.~Mehran~Mirramezani and Natalia~Rubio for insightful and helpful discussions and Jakob Richter for code improvements in \texttt{svZeroDSolver}. This work was supported by NIH Grants R01LM013120 and R01EB029362. The authors gratefully acknowledge the Stanford Research Computing Center for providing the computational resources necessary to the numerical simulations presented in this work.

\appendix
\section{Numerical solution in svZeroDSolver}
\label{sec_appendix_0d}
\begin{figure}[Htp!]
\centering
\includegraphics[width=5cm]{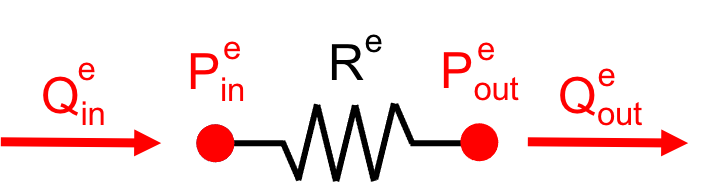}
\caption{A Poiseuille-based linear resistor lumped-parameter element. The solution variables for this element are shown in red.\label{fig_resistor}}
\end{figure}

In this appendix, we outline the spatial and temporal discretization and iterative solution process in \texttt{svZeroDSolver}. The 0D model is governed by combinations of lumped-parameter elements that can be combined to create a general nonlinear system of equations,
\begin{align}
\vc{E}\left(\vc{y}, t\right)\cdot\dot{\vc{y}} + \vc{F}\left(\vc{y}, t\right)\cdot \vc{y} + \vc{c}\left(\vc{y}, t\right) = \vc{0}.
\label{eqn_0d_dae_appendix}
\end{align}
Here, $\vc{y}$ is the global vector of solution variables (pressure and flow at LPN nodes), $\vc{E}$ and $\vc{F}$ are their associated \revb{coefficient functions}, and $\vc{c}$ is a vector of constants.\cite{verma20} For simplicity of notation, we drop the dependence on the solution vector and time in the following. We leverage the inherent modular nature of 0D models to create our \texttt{svZeroDSolver} software. The governing equations for a single element, $e$, can also be cast into the form of
\begin{align}
\textbf{E}^{e} \cdot\dot{\textbf{y}}^{e} + \textbf{F}^{e} \cdot\textbf{y}^{e} + \textbf{c}^{e} = \textbf{0}.
\end{align}
The local element contributions to $\textbf{E}^{e}$, $\textbf{F}^{e}$, and $\textbf{c}^{e}$, for each lumped-parameter element are defined in \texttt{svZeroDSolver} and assembled automatically into the global arrays,
\begin{align}
\textbf{E} = \underset{e = 1}{\overset{N_\text{elem}}{\mathbb{A}}}\textbf{E}^{e}, \quad
\textbf{F} = \underset{e = 1}{\overset{N_\text{elem}}{\mathbb{A}}}\textbf{F}^{e}, \quad
\textbf{c} = \underset{e = 1}{\overset{N_\text{elem}}{\mathbb{A}}}\textbf{c}^{e},
\end{align}
where $\mathbb{A}$ is the assembly operator and $N_\text{elem}$ is the total number of lumped-parameter elements in the 0D model. As an example, the governing equations for the Poiseuille-based resistor (Figure~\ref{fig_resistor}) are
\begin{align}
P_\text{in}^{e} - P_\text{out}^{e} - R^{e}Q_\text{in}^{e} &= 0,\\
Q_\text{in}^{e} - Q_\text{out}^{e} &= 0,
\end{align}
yielding the solution vectors and local element arrays
\begin{gather}
    \textbf{y}^{e} =
        \begin{bmatrix}
            P_\text{in}^{e} & Q_\text{in}^{e} & P_\text{out}^{e} & Q_\text{out}^{e}
        \end{bmatrix}^T, \quad
    \textbf{F}^{e} =
        \begin{bmatrix}
            1 & -R^{e} & -1 &  0 \ \cr
            0 &  1 &  0 & -1
        \end{bmatrix}, \quad
    \textbf{E}^{e} =
        \begin{bmatrix}
            0 & 0 & 0 &  0 \ \cr
            0 & 0 &  0 & 0
        \end{bmatrix}, \quad
    \textbf{c}^{e} =
        \begin{bmatrix}
            0 \ \cr
            0
        \end{bmatrix}.
\end{gather}
Observe that $\textbf{E}^{e}$ and $\textbf{c}^{e}$ are zero, due to the lack of differential terms and constant terms in the resistor's local governing equations.


After assembling our full 0D model, we solve Equation~\eqref{eqn_0d_dae_appendix} using the implicit generalized-$\alpha$ method.\cite{jansen00} The generalized-$\alpha$ method is an implicit predictor-corrector method that requires a single predictor step and a series of multi-corrector steps to solve for the solutions at each time step. Similar to other predictor-corrector schemes, the solutions are evaluated at intermediate times between $t_{n}$ and $t_{n + 1}$. However, in the generalized-$\alpha$ method, $\textbf{y}$ and $\dot{\textbf{y}}$ are evaluated at different intermediate times: 
%
\begin{align}
\textbf{y}_{n+\alpha_{f}} &= \textbf{y} \left( t_{n} + \alpha_{f}\Delta t \right),\\
\dot{\textbf{y}}_{n+\alpha_{m}} &= \dot{\textbf{y}} \left( t_{n} + \alpha_{m}\Delta t \right).
\end{align}
Here, $\alpha_{m}$ and $\alpha_{f}$ are the generalized-$\alpha$ parameters, where $\alpha_{m} = \frac{3 - \rho}{2 + 2\rho}$, $\alpha_{f} = \frac{1}{1 + \rho}$, and $\rho$ is the spectral radius. For each time step, the solution strategy is outlined below.

\begin{enumerate}
    \item $\textbf{Predictor step}$: Make an initial guess for $\textbf{y}_{n+1}$ and $\dot{\textbf{y}}_{n+1}$,
        \begin{align}
        \textbf{y}_{n+1} &= \textbf{y}_{n},\\
        \dot{\textbf{y}}_{n+1} &= \frac{\gamma - 1}{\gamma}\dot{\textbf{y}}_{n}, \quad\text{with~}\gamma = 0.5 + \alpha_{m} - \alpha_{f}
        \end{align}

    \item $\textbf{Initiator step}$: Interpolate $\textbf{y}_{n+\alpha_{f}}$ and $\dot{\textbf{y}}_{n+\alpha_{m}}$,
        \begin{align}
        \textbf{y}_{n+\alpha_{f}}^{k=0} &= \textbf{y}_{n} + \alpha_{f}\left(\textbf{y}_{n+1} - \textbf{y}_{n}\right),\\
        \dot{\textbf{y}}_{n+\alpha_{m}}^{k=0} &= \dot{\textbf{y}}_{n} + \alpha_{m}\left(\dot{\textbf{y}}_{n+1} - \dot{\textbf{y}}_{n}\right)
        \end{align}
    \item $\textbf{Multi-corrector step}$: Iteratively update the guess of $\dot{\textbf{y}}_{n+\alpha_{m}}^{k}$ and $\textbf{y}_{n+\alpha_{f}}^{k}$ for iteration $k$.

        We desire the residual, $\textbf{r}$, to be zero, where
        \begin{equation}
        \textbf{r}^k \vcentcolon= \textbf{E}^k \cdot\dot{\textbf{y}}_{n+\alpha_{m}}^{k} + \textbf{F}^k \cdot\textbf{y}_{n+\alpha_{f}}^{k}
         + \textbf{c}^k \stackrel{!}{=} \textbf{0}, \quad\text{with~} (\bullet)^k = (\bullet)\left(\dot{\textbf{y}}_{n+\alpha_{m}}^{k}, \textbf{y}_{n+\alpha_{f}}^{k}, t_{n+\alpha_{f}}\right).
        \end{equation}
        Using the Newton-Raphson method, we linearize this equation about $\textbf{y}_{n+\alpha_{f}}^{k}$ to obtain
        \begin{align}
        \textbf{K}^k \cdot\Delta \textbf{y}_{n+\alpha_{f}}^{k} = -\textbf{r}^k,
         \label{eq_newton_update}
        \end{align}
        where $\textbf{K}^k$ is the tangent matrix, which is calculated as
        \begin{equation}
        \begin{aligned}
        \textbf{K}^k = \frac{\partial \textbf{r}^k}{\partial \textbf{y}_{n+\alpha_{f}}} = \underset{\textbf{dE}}{\underbrace{\frac{\partial \textbf{E}^k}{\partial \textbf{y}_{n+\alpha_{f}}}\cdot\dot{\textbf{y}}_{n+\alpha_{m}}}} + \frac{\alpha_{m}}{\alpha_{f}\gamma\Delta t}\textbf{E}^k
         + \underset{\textbf{dF}}{\underbrace{\frac{\partial \textbf{F}^k}{\partial \textbf{y}_{n+\alpha_{f}}}\cdot\textbf{y}_{n+\alpha_{f}}}} + \textbf{F}^k
         + \underset{\textbf{dc}}{\underbrace{\frac{\partial \textbf{c}^k}{\partial \textbf{y}_{n+\alpha_{f}}}}}.
        \end{aligned}
        \end{equation} 
        As for to the local contributions $\textbf{E}^e$, $\textbf{F}^e$, and $\textbf{c}^e$, the local tangents $\textbf{dE}^e$, $\textbf{dF}^e$, and $\textbf{dc}^e$ must be defined for each 0D element. Note that in the linear resistor example, all tangents are zero.
    
        We solve \eqref{eq_newton_update} for $\Delta \textbf{y}_{n+\alpha_{f}}^{k}$ using a sparse direct solver and update the solution in this Newton-Raphson step
        \begin{align}
        \textbf{y}_{n+\alpha_{f}}^{k+1} &= \textbf{y}_{n+\alpha_{f}}^{k} + \Delta \textbf{y}_{n+\alpha_{f}}^{k},\\
        \dot{\textbf{y}}_{n+\alpha_{m}}^{k+1} &= \dot{\textbf{y}}_{n+\alpha_{m}}^{k} + \frac{\alpha_{m}}{\alpha_{f}\gamma\Delta t}\Delta \textbf{y}_{n+\alpha_{f}}^{k},
        \end{align}
        until the residual is lower than a given tolerance in iteration $k_\text{tol}$.
        
    \item $\textbf{Update step}$: Update solution and its time derivative in this time step,
        \begin{align}
        \textbf{y}_{n+1} &= \textbf{y}_{n} + \frac{\textbf{y}_{n+\alpha_{f}}^{k_\text{tol}} - \textbf{y}_{n}}{\alpha_{f}},\\
        \dot{\textbf{y}}_{n+1} &= \dot{\textbf{y}}_{n} + \frac{\dot{\textbf{y}}_{n+\alpha_{m}}^{k_\text{tol}} - \dot{\textbf{y}}_{n}}{\alpha_{m}}.
        \end{align}
\end{enumerate}


\newpage

\bibliography{references}

\end{document}